\documentclass[nologo,a4paper,10pt]{ETHpaper}
\usepackage[british]{babel}
\usepackage{blkarray} % for matrix
\usepackage[square,numbers]{natbib}
\usepackage{bbold}
\usepackage{xr-hyper}
\usepackage[colorlinks=true, linkcolor=black, citecolor=blue]{hyperref}
\usepackage[capitalise]{cleveref}

\def\multiset#1#2{\ensuremath{\left(\kern-.4em\left(\genfrac{}{}{0pt}{}{#1}{#2}\right)\kern-.4em\right)}}

\title{Empirical Networks are Sparse:\\ Enhancing Multi-Edge Models with Zero-Inflation}
\titlealternative{Empirical Networks are Sparse: Enhancing Multi-Edge Models with Zero-Inflation}
\renewcommand*{\thefootnote}{\fnsymbol{footnote}}
\author{Giona Casiraghi\footnote{Corresponding author, \href{mailto:gcasiraghi@ethz.ch}{gcasiraghi@ethz.ch}}, Georges Andres}
\authoralternative{Giona Casiraghi, Georges Andres}

\address{Chair of Systems Design, ETH Z\"urich, Weinbergstrasse 56/58, 8092 Z\"urich, Switzerland \\[2mm] }
\www{\url{http://www.sg.ethz.ch}}

\begin{document}
\maketitle
\renewcommand*{\thefootnote}{\arabic{footnote}}

\begin{abstract}
Real-world networks are sparse.
As we show in this article, even when a large number of interactions is observed, most node pairs remain disconnected.
We demonstrate that classical multi-edge network models, such as the \(G(N,p)\), configuration models, and stochastic block models, fail to accurately capture this phenomenon.
To mitigate this issue, zero-inflation must be integrated into these traditional models.
Through zero-inflation, we incorporate a mechanism that accounts for the excess number of zeroes (disconnected pairs) observed in empirical data.
By performing an analysis on all the datasets from the Sociopatterns repository, we illustrate how zero-inflated models more accurately reflect the sparsity and heavy-tailed edge count distributions observed in empirical data.
Our findings underscore that failing to account for these ubiquitous properties in real-world networks inadvertently leads to biased models that do not accurately represent complex systems and their dynamics.

\end{abstract}

\section{Introduction}\label{sec:introduction}

Networks are foundational for understanding complex systems.
Accurate modelling of these networks can significantly impact various domains, such as optimising distribution systems~\cite{amico2024}, understanding the spread of diseases~\cite{disease_spread}, and analysing social behaviours~\cite{social_behaviors}.
However, a critical challenge lies in developing models that correctly capture the complex patterns observed in real-world networks.

Consider the interactions between students in a high school.
Interactions can be represented as edges in a network where students are the nodes.
Pairs of students who do not interact appear as disconnected, while repeated interactions between the same pair of students create multi-edges (also known as parallel edges), where the multiplicity of each edge represents the interaction count.
Students from different classes have limited opportunities to meet due to distinct schedules and social circles.
When they don’t share common spaces or schedules, they don't interact.
However, when they do meet, the frequency of their interactions can vary independently of their initial chance of meeting.

\begin{figure}[t]
  \centering
\includegraphics[width=.8\textwidth]{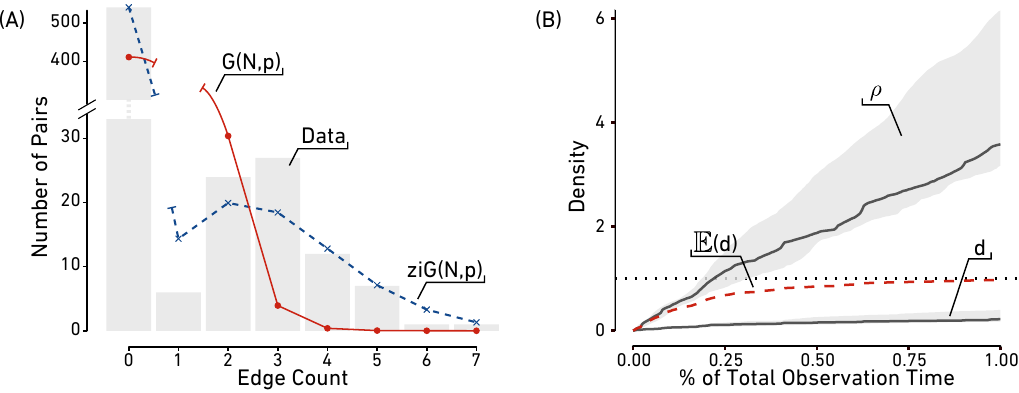}
  \caption{Empirical multi-edge networks are sparse. Traditional multi-edge models like the G(N,p) struggle to reflect real-world data characteristics. 
  \textbf{(A)} Edge count distribution in Zachary's Karate Club showing bimodality. The red solid line represents the G(N,p) prediction, and the blue dashed line its zero-inflated counterpart. 
  \textbf{(B)} Over time, edges accumulate between the same pairs of nodes in real-world networks. In grey, the interquartile range of the number of multi-edges per pair \( \rho = m/\binom{N}{2} \) and the fraction of connected node pairs \( d = M/\binom{N}{2} \), over all Sociopatterns datasets. The black lines denote the median values, while the dashed red line represents the expected fraction of connected pairs according to the G(N,p) model with the corresponding \( m \) value. Note that while the model quickly predicts a fully connected network, the empirical network remains sparse, indicating that most interactions occur among existing pairs rather than forming new connections.}
  \label{fig:illustration}
\end{figure}

This dynamic, common in real-world networks, often leads to ``zero-inflation'' in the distribution of interaction counts:
the number of disconnected pairs exceeds what we would expect given the large number of potential interactions.
This results in an inflated number of zeroes in the distribution of interaction counts, as illustrated in \cref{fig:illustration}~(A).

In other words, empirical networks are typically sparse~\cite{faloutsos2004connection,Decelle2011,Krivitsky2011}:
only a limited number of node pairs share multiple interactions, while the majority remains disconnected.
Traditional network models, such as the \(G(N,p)\)~\cite{gnp,gilbert}, configuration models~\cite{chung-lu, fosdick, casiraghiurn}, and stochastic block models~\cite{karrernewman, Peixoto2017}, have been instrumental in advancing our understanding of complex networks.
However, these models fall short in representing the inherent sparsity observed in real multi-edge network data.
They usually assume a proportional relationship between the growth of edges and the number of connected node pairs, which is not always accurate.

Our article emphasises the necessity of incorporating \emph{zero-inflation}~\cite{Lambert1992} into network models.
To bridge this gap, we systematically extend a broad family of multi-edge network models to effectively address the challenges posed by sparse empirical networks.

Recently, there has been an increased focus on the issue of zero-inflation in network data.
\citet{Krivitsky2012} briefly discussed a zero-inflated version of exponential random graph models for multi-edge networks that accounts for sparsity by modelling dyad-wise distributions.
\citet{choi2023model} highlighted the challenges posed by zero-inflation when modelling sparse networks and proposed methods to address this issue.
Similarly, \citet{Ebrahimi2021} and \citet{Motalebi2021} explored zero-inflated and hurdle models to better capture the inherent sparsity in social and biological networks.
Furthermore, \citet{Dong2019} and \citet{Motalebi2021dcsbm} specifically focused on adapting stochastic block models to account for excess zeroes, underscoring the importance of accurately modelling sparsity for realistic network analysis.
Collectively, these works emphasise the necessity of incorporating zero-inflation into network models to enhance their applicability to real-world, sparse network data.

This work aims to achieve two main objectives.
First, we highlight the ubiquity of zero-inflation in real-world multi-edge network data by analysing \emph{all} datasets from the Sociopatterns repository~\cite{sociopatterns}.
Second, we demonstrate how zero-inflation can be integrated into traditional multi-edge network models, providing a more accurate representation of the sparse nature of empirical networks.
Adopting zero-inflated models in network science holds promise for improving the analysis of intrinsically sparse structures, such as higher-order networks~\cite{mogen, ingokdd} and hypergraphs~\cite{hypergraphs}.

\subsection{Multi-edge network models}\label{sec:general_idea}

Multi-edge network models serve as generative stochastic frameworks that not only capture the \emph{presence} of edges between nodes but also quantify the \emph{number} of edges observed for each node pair.
These models are crucial for understanding the complex structures and dynamics of real-world networks.

Broadly, multi-edge network models fall into three main categories: micro-canonical~\cite{Peixoto2013,fosdick}, Poisson~\cite{Norros2006,karrernewman}, and Hypergeometric~\cite{casiraghiurn,bccm}.
Each category offers a different level of constraint and flexibility, catering to various types of network data.

Micro-canonical models are the most constrained, defining an ensemble of networks that share specific \emph{fixed} attributes, such as degree sequences or intra-block edge counts, and assigning equal probability to each network within the ensemble.
In contrast, Poisson and Hypergeometric models operate within more flexible sample spaces, preserving these attributes only \emph{in expectation}.
The critical difference between Poisson and Hypergeometric models lies in their independence assumptions: the former treat node pairs as independent, whereas the latter do not~\cite{bccm}.
This distinction is crucial, as it influences the applicability and performance of the models on different types of network data.

In this article, we focus on Poisson network models.
The independence assumption in Poisson models simplifies the introduction of zero-inflation, which is our primary aim.

Poisson multi-edge network models rely on Poisson count processes to describe phenomena such as edge formation and node interactions.
This is the case for classic extensions of the \(G(N,p)\) model to multi-edge networks~\cite{gnp,gilbert}, the Chung-Lu configuration model~\cite{chung-lu,Norros2006}, the classical stochastic block model and the degree-corrected stochastic block model~\cite{karrernewman}, and even count-ERGM models~\cite{Krivitsky2012}.

In a Poisson count process, the probability of observing \( n \) events in a fixed time interval is governed by a Poisson distribution with parameter \( \lambda \):
\begin{equation}
  \Pr(X=n) = \frac{\lambda^n}{n!} e^{-\lambda} \, .
\end{equation}
This Poisson model assumes that the distribution of the random variable \( X \) will be centred around its mean \( \lambda \), with a low probability of zero occurrences, especially as \( \lambda \) increases.

Poisson network models are defined as \(n^2\) independent Poisson count processes, one for every (directed) pair of nodes in the network:
\begin{equation}\label{eq:poisson_network}
  \Pr(\mathcal{G}) = \prod_{ij} \Pr(A_{ij}) = \prod_{ij} \frac{\lambda_{ij}^{A_{ij}}}{A_{ij}!}\exp(-\lambda_{ij}),
\end{equation}
where \( A_{ij} \) denotes the edge count between $i$ and $j$ and \( \lambda_{ij} \) are parameters that can be functions of node attributes, edge attributes, or block memberships, among others.
Appropriately specifying these parameters is key for capturing the heterogeneous nature of interactions within the network.

Several well-known network models are special cases of this general Poisson framework.
For example, the \( G(N,p) \) model for multi-edge networks corresponds to a Poisson network model where \( \lambda_{ij} = p \) for all \( i, j \).
The Stochastic Block Model is a generalisation that introduces \( B \) different \( \lambda_b \) parameters, one for each block in the network~\cite{karrernewman}.
The Chung-Lu Configuration Model, foundational to many concepts in network science like modularity~\cite{girvan} and degree-correction~\cite{karrernewman}, also fits within this framework, featuring node-dependent parameters \( \lambda_{ij} = \theta_i \theta_j \)~\cite{chung-lu,Norros2006}.
Further, the Degree-Corrected Stochastic Block Model combines aspects of both the Chung-Lu and Stochastic Block Models, with \( \lambda_{ij} = \theta_i \theta_j \lambda_b \)~\cite{karrernewman}.

\subsection{Modelling zero-inflation}

While Poisson network models are standard tools in multiple disciplines~\cite{karrernewman,Krivitsky2012,Norros2006}, they struggle to accurately represent sparse multi-edge networks~\cite{Lambert1992,choi2023model,Dong2019,Motalebi2021,Ebrahimi2021,Motalebi2021dcsbm}.
The expected number of connected pairs \(M\) according to a Poisson network model is given by:
\begin{equation}\label{eq:zero}
  \mathbb{E}[M] = \sum_{ij} \left(1 - \exp\left(-\lambda_{ij}\right)\right).
\end{equation}
However, \(\sum_{ij} \lambda_{ij}\) gives the expected number of multi-edges \(m\) in the network.
Consequently, \(\mathbb{E}[M]\) saturates exponentially with \(\mathbb{E}[m]\) to a fully-connected network.
Hence, with a larger observed \(m\), the model tends to yield fully-connected networks.

\Cref{fig:illustration}~(B) compares the sparsity of empirical multi-edge networks with the predictions of Poisson network models.
We use interaction networks from the \textit{Sociopatterns} repository as an example~\cite{sociopatterns} , where different datasets record contacts between individuals over short time frames.
With increasing observation time and data collection, the number of multi-edges \(m\) grows rapidly.
Yet, the growth of the number of connected node pairs \(M\) is considerably slower.
This suggests that individuals predominantly interact within their existing social circles, limiting the number of new interactions over short periods.
The dashed line in the plot shows the predicted number of connected pairs according to \cref{eq:zero}.
As discussed above, with an increasing number of multi-edges \(m\), the number of connected pairs quickly saturates to a fully connected network, deviating from the empirical data.

The observed \emph{count distribution}, instead, shows a strong bimodality, characterised by peaks at zero and around some \( \hat{\lambda} \) value.
Even the well-known example of Zachary's Karate Club network exhibits this bimodal distribution, as illustrated in \cref{fig:illustration}~(A).
Zachary's Karate Club represents the social interactions within a karate club.
When examining the distribution of interactions in this network, we observe a significant number of disconnected node pairs, even when considering the existence of smaller groups, exemplifying the zero-inflation phenomenon.

Zero-inflated models~\cite{Lambert1992} have been developed to mitigate this issue.
These models are a mixture of a binary process for generating zeros and a Poisson count process for generating the counts.
The probability mass function is given by:
\begin{equation}\label{eq:zi-equation}
  \Pr(X=n) = (1-q) \delta_0(n) + q \frac{\lambda^n}{n!} e^{-\lambda} \, .
\end{equation}
Here, \( q \in[0,1]\) is the mixture weight, and \( \delta_0(n) \) is the Kronecker delta function, which is 1 when \( n=0 \) and 0 otherwise.
The term \( (1-q) \) accounts for the excess zeros that are not explained by the Poisson process, providing a more accurate model for data like the one in \cref{fig:illustration}~(A).

In addition to zero-inflated models, \emph{hurdle models} provide another approach to address count data with excess zeroes~\cite{Feng2021}.
While zero-inflated models combine a binary process for zero counts with a Poisson process for positive counts, hurdle models treat zeros and positive counts as outcomes of two distinct processes.
Specifically, a hurdle model first uses a binary process to determine whether an interaction occurs (i.e., whether the count is zero or positive), and then applies a truncated Poisson process for positive counts.
This separation ensures that zeros are generated differently than positive counts, which can offer advantages in terms of model identifiability and interpretability~\cite{Motalebi2021dcsbm}.
However, this assumption implies that sparsity is \emph{only} generated by the hurdle process, and not by low interaction rates.
Such a strict assumption may not always hold true, especially in cases where modelling the interplay between low interaction rates and zero-inflation is critical~\cite{Feng2021}.
Finally, if the hurdle is set at zero, as most commonly done, it can be shown that the resulting hurdle model is just a re-parametrized zero-inflated model~\cite{johnson2005}.

In this article, we choose to focus only on zero-inflation as a way to model sparsity in complex networks.
In the following section, we detail how zero-inflation is incorporated in Poisson network models and how to perform the inference of the parameters from data.
Nevertheless, most of the results shown apply in a similar way to hurdle models~\cite{Feng2021}.

\paragraph{Parameter Estimation}
Throughout this article, we employ a variation of the Expectation-Maximization (EM) algorithm for parameter inference, specifically tailored to leverage key properties of zero-inflated Poisson (ZIP) distributions.
This approach ensures methodological consistency across different models, allowing for coherent comparisons and a clearer understanding of their individual characteristics.

Direct Maximum Likelihood Estimation (MLE) for zero-inflated models presents significant challenges, as noted in prior work~\cite{Sari2021, Ali2022, Consul1992}.
The difficulty stems from the complexity of network models in high-dimensional spaces, where the MLE of the mixture probability in \cref{eq:zi-equation} involves a logarithm of sums.
This form complicates simplification and resists manipulation with standard tools, making numerical optimization of the log-likelihood function necessary.

Rather than directly estimating the parameters \( q \) and \( \lambda \) in \cref{eq:zi-equation}, we leverage the fact that the ZIP distribution is a power series distribution (PSD).
As shown in~\cite{johnson2005}, PSDs allow us to use the first moment equation as a maximum-likelihood equation.
Solving
\begin{equation}\label{eq:m1}
\bar{x} = \mathbb E[n] = \widehat{\lambda} \widehat{q}
\end{equation}
yields one of the maximum-likelihood equations for parameter estimation.

In the univariate case, the second MLE equation is defined by matching the observed proportion of zeroes \( f_0 \) to its expected value~\cite{johnson2005}:
\begin{equation}\label{eq:m2}
(1 - \widehat{q}) + \widehat{q} e^{-\widehat{\lambda}} = f_0.
\end{equation}

In the univariate case of \cref{eq:zi-equation}, fixing the first two moments--one for the count process and one for the zero process--effectively achieves the MLE of the parameters.
For the more complex models discussed below, however, the second equation may not always be applicable due to dependencies between parameters.
In network models with structured dependencies (e.g., degree-corrected models), the mixture parameter \( q \) may vary by node pair or community, introducing correlations that complicate isolating the zero-inflation component.

To address this, we rely on the first equation, which remains valid as an MLE condition, to link the Poisson and mixture parameters.
We then treat the Poisson parameters as latent and optimize over the mixture ones, effectively defining a one-step EM algorithm.

\section{Zero-Inflating Poisson Network Models}\label{sec:zero_inflation}
\subsection[Zero-Inflated G(N,p) Model]{Zero-Inflated \(G(N, p)\) Model (zi-\(G(N, p)\))}\label{sec:zignp}

The \(G(N, p)\) model is one of the foundational generative models for networks and among the simplest.
It is characterised solely by a parameter \(p\), which determines the expected number of edges in a network realisation~\cite{gilbert}.
In this model, interactions between different node pairs are assumed to be independent and identically distributed.

For the multi-edge variant of the \(G(N, p)\) model, we set \(\lambda_{ij} = p\) in \cref{eq:poisson_network} for all node pairs.
The expected number of connected node pairs---henceforth denoted as \emph{links}---in a network realisation is then given by:
\begin{equation}
  \mathbb{E}(M|p) = \sum_{ij} (1 - e^{-p}) = N^2 - N^2 e^{-p}\,,
\end{equation}
where \(N\) represents the number of nodes, considering a loopy directed network.
This model reveals its limitation in representing sparse multi-edge networks, as \(\mathbb{E}(M|p)\) approaches \(N^2\) with increasing \(p\).

To better model sparse networks, we integrate zero-inflation into the edge probabilities using \cref{eq:zi-equation}.
The zero-inflated \(G(N, p)\) model is defined by:
\begin{equation}\label{eq:zignp}
  \Pr(\mathcal{G}|p, q) = \prod_{ij} \left( (1-q)\delta_0(A_{ij}) + q\frac{p^{A_{ij}}}{A_{ij}!}e^{-p} \right).
\end{equation}

Incorporating zero-inflation, we calculate the expected number of interactions (edges) \(\mathbb{E}(m|p, q)\) and the expected number of links \(\mathbb{E}(M|p, q)\) as follows:
\begin{align}
  \mathbb{E}(m|p, q) &= q p N^2\,, \\
  \mathbb{E}(M|p, q) &= q N^2 - N^2 q e^{-p}\,.
\end{align}

As the zi-\(G(N, p)\) model corresponds to the univariate ZIP distribution discussed above, the maximum likelihood estimates of the parameters \(p\) and \(q\) can be obtained by matching these expected values to the observed counts of interactions and links, \(\widehat{m}\) and \(\widehat{M}\), respectively:
\begin{equation}
  \mathbb{E}(m|p, q) := \widehat{m}\,, \qquad \mathbb{E}(M|p, q) := \widehat{M}\,.
\end{equation}
Solving for \(p\) and \(q\), we derive:
\begin{equation}
  \widehat{p} = \frac{\widehat{m}}{N^2 \widehat q}\,, \qquad
  \widehat{q} = \frac{\widehat{m} \widehat{M}}{N^2 (\widehat{m} + \widehat{M} \mathcal{W}\left[-\frac{\widehat{m} e^{-\frac{\widehat{m}}{\widehat{M}}}}{\widehat{M}}\right])}\,,
\end{equation}
where \(\mathcal{W}[z]\) represents Lambert's W function, solving \(w e^w = z\)~\cite{Lambert1992}.

\subsection{Zero-Inflated Stochastic Block Model (zi-SBM)}\label{sec:zisbm}

Building on the \(G(N, p)\) model, the Stochastic Block Model (SBM) introduces a richer representation of network structures by distinguishing blocks in the network, each characterised by a unique edge probability~\cite{Holland1983,aicher2015learning}.
By doing so, the model captures community structures.
Nodes are assigned to one of \(B\) different groups, and the probability of interactions between nodes \(i\) and \(j\) in groups \(b_i\) and \(b_j\), respectively, is determined by a block-specific parameter \(\lambda_{b_i b_j}\).
Hereafter, we assume block assignments to be known a priori.
We discuss the implications for community detection of zero-inflation later in the manuscript.

To accommodate potential sparsity of interactions within different blocks, we detail the zero-inflated version of the SBM, denoted as zi-SBM.
The model is defined by the following probability distribution:
\begin{align}\label{eq:zisbm}
  \begin{split}
    \Pr&(\mathcal{G}|\pmb{\lambda}, \pmb{q}) = \\ &\prod_{ij} \left( (1-q_{b_i b_j})\delta_0(A_{ij}) + q_{b_i b_j}\frac{\lambda_{b_i b_j}^{A_{ij}}}{A_{ij}!}\exp(-\lambda_{b_i b_j}) \right),
  \end{split}
\end{align}
where each block \((b, d)\) is associated with a unique mixture parameter \(q_{bd}\), allowing for block-specific zero-inflation.

The expected number of interactions \(m_{bd}\) and links \(M_{bd}\) in the zi-SBM are given by:
\begin{align}
  &\mathbb{E}(m_{bd}|\pmb{\lambda}, \pmb{q}) = q_{bd}\lambda_{bd}N_b N_d, \\
  &\mathbb{E}(M_{bd}|\pmb{\lambda}, \pmb{q}) = \sum_{bd} N_b N_d q_{bd} (1 - e^{-\lambda_{bd}}),
\end{align}
where \(N_b\) denotes the number of nodes in group \(b\).

For parameter inference, we equate the first moments of the distribution to the observed values \(\widehat{m}_{bd}\) and \(\widehat{M}_{bd}\):
\begin{equation}\label{eq:sbm_infer}
  \mathbb{E}(m_{bd}|\lambda_{bd}, q_{bd}) := \widehat{m}_{bd}, \qquad \mathbb{E}(M_{bd}|\lambda_{bd}, q_{bd}) := \widehat{M}_{bd}.
\end{equation}
Solving \cref{eq:sbm_infer}, we find:
\begin{equation}
  \widehat{\lambda}_{bd} = \frac{\widehat{m}_{bd}}{q_{bd} N_b N_d}.
\end{equation}
A closed-form solution for \(\widehat{q}_{bd}\) is not readily available.
Nonetheless, the values of \(\widehat{q}_{bd}\) can be determined by numerically solving the set of \(B^2\) independent equations given in \cref{eq:sbm_infer}.
Because the zi-SBM effectively defines B independent zi-\(G(N, p)\) models, i.e., one per block, $\widehat\lambda_{bd}$ and \(\widehat{q}_{bd}\) are the maximum likelihood estimates for the model parameters.

\citet{Dong2019} proposed a variational-EM algorithm for parameter estimation in the special case of a multilayer zero-inflated stochastic block model.
This algorithm efficiently estimates the community labels and model parameters, handling the sparsity and correlations in multilayer networks.
The method proposed demonstrates effectiveness in capturing complex interaction patterns through extensive simulations and real-world case studies.

\subsection{Zero-Inflated Configuration Model (zi-CLCM)}\label{sec:zi-CLCM}

While both the \(G(N,p)\) and SBM models offer valuable insights, they fall short in encoding node heterogeneities.
Configuration models fill this gap by introducing a parameterisation that accounts for degree heterogeneities~\cite{fosdick}.
In the framework of Poisson models, the Chung-Lu configuration model (CLCM) achieves this by expressing the general parameters \(\lambda_{ij}\) as \(\theta^\text{out}_i \theta^\text{in}_j\), a product of node-parameters~\cite{chung-lu,Norros2006}.
For undirected networks, we have \(\pmb{\theta^\text{out}}=\pmb{\theta^\text{in}}=\pmb\theta\).

Incorporating zero-inflation into the CLCM results in the Zero-Inflated Chung-Lu Configuration Model (zi-CLCM).
The model is described by the probability distribution:
\begin{align}\label{eq:ziclcm}
    \begin{split}
      \Pr&(\mathcal{G}|\pmb\theta,q) =\\ &\prod_{ij} \left( (1-q)\,\delta_0(A_{ij}) + q\,\frac{(\theta^\text{out}_i \theta^\text{in}_j)^{A_{ij}}}{A_{ij}!}\exp(-\theta^\text{out}_i \theta^\text{in}_j) \right).
    \end{split}
\end{align}

\paragraph{Parameters Identifiability}
It is important to note that the model's Poisson parameters \(\pmb{\theta^\text{out}}\) and \(\pmb{\theta^\text{in}}\) appear always in pairs.
This means that they are defined modulo constants.
This characteristic requires fixing a constraint to enable their estimation.
A common approach is to re-parametrize the model by introducing a new parameter \(C\) that constrains the L1-norm of \(\pmb{\theta^\text{out}}\) and \(\pmb{\theta^\text{in}}\):
% Exploiting the relation between \cref{eq:m1} and the maximum likelihood equation, we get that the constraint for the L1-norm of the parameter vectors \(\pmb{\theta^\text{out}}\) and \(\pmb{\theta^\text{in}}\) given by
%
\begin{equation}\label{eq:clcm-constraint}
  \sum_{i}\theta^\star_i = C,
\end{equation}
where the superscript $^{*}$ stands for in/out.

% \begin{equation}\label{eq:clcm-constraint}
%   \sum_{i}\theta^\star_i = \sqrt{\frac{\widehat{m}}{q}}
% \end{equation}
%
% yields the maximum likelihood solution.
% This constraint ties the expected number of interactions in the network to the node-specific parameters, allowing their inference.

\paragraph{Parameters Estimation}
To define the first set of maximum likelihood equations, corresponding to \cref{eq:m1} for the univariate case, we set three different first-moment equations to their observed values:
\begin{align}
  \mathbb{E}(m|\pmb\theta,q) := \widehat{m}, \label{eq: clcm-m}\\
  \forall i:\; \mathbb{E}(k^\text{out}_i|\pmb\theta,q) := \widehat{k^\text{out}_i}, &\quad \forall i:\;\mathbb{E}(k^\text{in}_i|\pmb\theta,q) := \widehat{k^\text{in}_i},
\end{align}
where \(\widehat{m}\) denotes the number of interactions in an observed network and \(\widehat{k^\text{out}_i}\) and \(\widehat{k^\text{in}_i}\) the observed out- and in-degrees of $i$.

Replacing \cref{eq:clcm-constraint} in \cref{eq: clcm-m} yields
\begin{equation}
  \label{eq:1}
   \mathbb{E}(m|\pmb\theta,q) = q\sum_{i,j} \theta^{\text{out}}_{i} \theta^{\text{in}}_{j} = qC^{2},
 \end{equation}
which allows to tie the maximum-likelihood estimator $\widehat{C}$ to $q$:
\begin{equation}\label{eq:chat}
 \widehat{C} = \sqrt{\frac{\widehat{m}}{q}}\,.
\end{equation}
The expected degree sequences from the model are given by:
\begin{align}
    \begin{split}
      \mathbb{E}(k^\text{out}_i|\pmb\theta,q)= q\,\theta^\text{out}_i\sum_{j}\theta^\text{in}_j, \\ \mathbb{E}(k^\text{in}_i|\pmb\theta,q) = q\,\theta^\text{in}_i\sum_{j}\theta^\text{out}_j, \label{eq:clcm-deg-exp}
        \end{split}
\end{align}
From \cref{eq:clcm-constraint,eq:clcm-deg-exp,eq:chat}, we derive expressions for \(\widehat{\theta^\star_i}\) as functions of \(q\) and the observed degrees \(\widehat{k^\text{out}_i}\) and \(\widehat{k^\text{in}_i}\):
\begin{equation}\label{eq:momtheta}
  \widehat{\theta^\star_i} = \frac{\widehat{k^\star_i}}{\sqrt{\widehat{m} q}}\,.
\end{equation}
Lastly, we need to estimate the remaining parameter $q$.
In this case, however, we need to resort to the explicit maximization of the likelihood function.
Substituting \cref{eq:momtheta} into \cref{eq:ziclcm} leads to an equation for \(\widehat{q}\) that can be optimized numerically.

\subsection{Zero-Inflated Degree-Corrected Stochastic Block Model (zi-DCSBM)}\label{sec:zi-DCSBM}

Finally, we discuss the Zero-Inflated Degree Corrected Stochastic Block Model (zi-DCSBM).
As for the standard degree-corrected stochastic block model~\cite{karrernewman}, it incorporates both node heterogeneity and block structure into a single comprehensive model, combining the features of the zi-CLCM and the zi-SBM.
The introduction of block-specific mixture parameters \(q_{bd}\), constituting the vector \(\pmb{q}\), allows the model to account for varying levels of zero-inflation across different blocks.
By parametrising \cref{eq:poisson_network} with \(\lambda_{ij} = \theta^\text{out}_i \theta^\text{in}_j \lambda_{b_i b_j}\), we derive the following probability distribution:
\begin{align}\label{eq:zidcsbm}
    \begin{split}
  \Pr(\mathcal{G}|\pmb{\theta},\pmb{\lambda},\pmb{q}) &= \prod_{ij} \biggl( (1-q_{b_i b_j})\,\delta_0(A_{ij}) + \\
                                                      &q_{b_i b_j}\,\frac{(\theta^\text{out}_i \theta^\text{in}_j \lambda_{b_i b_j})^{A_{ij}}}{A_{ij}!}\exp(-\theta^\text{out}_i \theta^\text{in}_j \lambda_{b_i b_j}) \biggr) .
    \end{split}
\end{align}
\paragraph{Parameters Identifiability}
Just like in the CLCM and zi-CLCM, the parameters \(\pmb{\theta^\text{out}}\) and \(\pmb{\theta^\text{in}}\) in the DCSBM and zi-DCSBM are defined modulo constants.
We re-parametrize the model by deriving $2 B$ constraints for the L1-norm of the node-specific Poisson parameters:
\begin{equation}
  \label{eq:dcsbm-constraint}
  \sum_{i}\theta^\star_i\delta_b(b_i) = C^\star_b,
\end{equation}
for each block $b$.
In contrast to the CLCM and zi-CLCM, here, the zi-DCSBM does not depend on the choice of parameters $C^\star_b$, as shown in Supplementary Information E.
Therefore, we are free to set $C^\star_b:=C=1$ $\forall b$ to simplify the parameter estimation.

\paragraph{Parameters Estimation}

%To define such a constraint, we exploit again the relation between \cref{eq:m1} and the maximum likelihood equation.

Manipulating the first moment equation, we get the expected values of the number of interactions per block:
\begin{equation}
  \widehat m_{bd} = \mathbb{E}(m_{bd}|\pmb{\theta},\pmb{\lambda},\pmb{q}) = q_{bd}\lambda_{bd}\sum_{i \in b}\theta^\text{out}_i \sum_{j \in d}\theta^\text{in}_j. \label{eq:dcsbm-block-exp}
  \end{equation}
The constraints in~\cref{eq:dcsbm-constraint} allow us to tie the block Poisson parameters $\pmb\lambda$ to the mixture parameters $\pmb q$:
\begin{equation}\label{eq:lambdahat}
  \widehat{\lambda}_{bd} = \frac{\widehat{m}_{bd}}{q_{bd}}.
\end{equation}

From the first moment equation, we derive the expected degree sequences
  \begin{align}
      \begin{split}
  \mathbb{E}(k^\text{out}_i|\pmb{\theta},\pmb{q}) = \theta^\text{out}_i \sum_{d} q_{b_i d} \lambda_{b_i d} \sum_{j \in d} \theta^\text{in}_j, \\
        \mathbb{E}(k^\text{in}_i|\pmb{\theta},\pmb{q}) = \theta^\text{in}_i \sum_{d} q_{b_i d} \lambda_{b_i d} \sum_{j \in d} \theta^\text{out}_j. \label{eq:dcsbm-deg-exp}
      \end{split}
  \end{align}
Leveraging the constraints given by \cref{eq:dcsbm-constraint}, from \cref{eq:dcsbm-deg-exp}, we obtain:
\begin{equation}\label{eq:dcsbm-deg-exp2}
  \mathbb{E}(k^\star_i|\pmb{\theta},\pmb{q}) = \theta^\star_i \sum_{d} q_{b_i d} \lambda_{b_i d}.
\end{equation}

Solving the set of equations defined by \cref{eq:lambdahat,eq:dcsbm-deg-exp2} for \(\pmb{\theta^\text{out}}\) and \(\pmb{\theta^\text{in}}\) yields their MLE:
\begin{equation}
  \widehat{\theta^\star_{i}} = \frac{\widehat{k^\star_i}}{\widehat{\kappa}^\star_{b_i}},
\end{equation}
where \(\widehat\kappa^\text{out}_{b_i} = \sum_d \widehat m_{b_i d}\) and \(\widehat\kappa^\text{in}_{b_i} = \sum_d \widehat m_{d b_i}\) denote the out- and in-degree of block \(b_i\), respectively.
Finally, substituting these expressions into \cref{eq:zidcsbm} provides a set of $B^2$ independent equations for each \(q_{bd}\) that need to be optimized numerically to find the MLEs of all remaining parameters.

\Citet{Motalebi2021dcsbm} discussed the zi-DCSBM in comparison with a hurdle version of the DCSBM.
The probability distribution of the hurdle-DCSBM can be written as
\begin{align}
    \begin{split}
  &\Pr(\mathcal{G}|\pmb{\theta},\pmb{\lambda},\pmb{q}) = \\
  &\prod_{ij}
  \begin{cases}
  (1-q_{b_i b_j}) \quad\text{if } A_{ij} = 0\,,\\
  q_{b_i b_j}\,\frac{(\theta^\text{out}_i \theta^\text{in}_j \lambda_{b_i b_j})^{A_{ij}}\exp(-\theta^\text{out}_i \theta^\text{in}_j \lambda_{b_i b_j})}{A_{ij}! \left(1-\exp(-\theta^\text{out}_i \theta^\text{in}_j \lambda_{b_i b_j})\right)} \quad\text{if } A_{ij} > 0\,,
  \end{cases}
    \end{split}
\end{align}
where the role of the model parameters is the same as in the case of the zi-DCSBM.
The primary difference between these two models lies in their approach to handling sparsity: 
zero-inflation accounts for excess zeros by introducing a separate zero-generating process, 
while hurdle models treat zeros and positive counts as generated by two distinct processes.
Hurdle models offer advantages in terms of identifiability because they eliminate the ambiguity between zero counts and low interaction rates.
However, this means that sparsity, or disconnected pairs, is \emph{solely} due to the hurdle process and not from inherently low interaction rates, which may limit their appropriateness for many applications~\cite{Feng2021}.

\paragraph{Node Level Zero-Inflation}
The models described by \cref{eq:ziclcm,eq:zidcsbm} account for node heterogeneity only at the Poisson level.
At the zero-inflation level, they assume a uniform process with a single parameter $q_{b}$ per block $b$ (or one single parameter for the zi-CLCM as all nodes belong to the same block).
This means that the model will reproduce the expected degrees but not the number of neighbors per node.
To introduce degree heterogeneity at the mixture level as well, we can modify $q_{b}$ into $q_{b}q^\text{out}_i$ $q^\text{in}_j$.
The parameter estimation process follows the same procedure of the simpler models.
However, now we need to optimize the likelihood equation for the $2\cdot N + B$ parameters $\pmb q$.
This increment in computational complexity (discussed in Supplementary Information D), though, is offset by a model that better reproduces the empirical data.

\paragraph{Considerations about community detection}
So far, we have assumed that the partitioning of nodes into distinct groups is known.
However, community detection can also be performed endogenously within the zero-inflated model, assigning labels to the nodes based on interaction and link data.
Community detection methods can generally be divided into two categories: those based on quality functions~\cite{girvan} and those based on Maximum Likelihood Estimation (MLE) principles~\cite{Peixoto2017}.

The former approach, based on quality functions, can be readily adapted to zero-inflated models.
To do so, one require a suitable partition quality function that needs to be optimized to find optimal node partitionings.
A common example is \emph{modularity} optimization~\cite{girvan}, where the partition quality function \(Q\) is given by:
\begin{equation}\label{eq:girvanQ}
  Q = \frac{1}{m} \sum_{ij} \left[ A_{ij} - \frac{k^\text{out}_i k^\text{in}_j}{m} \right] \delta(c_i, c_j),
\end{equation}
where \(A_{ij}\) represents the adjacency matrix, \(k_i\) and \(k_j\) are the observed degrees of nodes \(i\) and \(j\), \(m\) is the total number of edges, \(c_i\) and \(c_j\) are the communities of nodes \(i\) and \(j\), and \(\delta\) is the Kronecker delta function.
The term \(\frac{k^\text{out}_i k^\text{in}_j}{m}\) in $Q$ denotes the \emph{expected number of interactions} between nodes $i$ and $j$ according to the a CLCM fitted to the empirical network defined by the adjacency matrix $A$~\cite{girvan}.

In the case of the zi-CLCM, the expected number of interactions is given by \(q \theta^\text{out}_i \theta^\text{in}_j\).
Substituting the MLE estimates of $\pmb\theta$ from \cref{eq:momtheta}, we get that expected number of interactions according to the zi-CLCM fitted to the empirical network $A$ is also given by \(\frac{k^\text{out}_i k^\text{in}_j}{m}\).
Thus, the partition quality function $Q$ in \cref{eq:girvanQ} is equally applicable to zero-inflated models, meaning that community detection via modularity optimization is unaffected by the presence of zero-inflation.

The latter category of community detection instead, exemplified by methods utilising information theory to assess community quality, requires the computation of the MLE of the model and potentially the integration of the likelihood function to eliminate continuous parameters like \(\theta^\star\) and \(q\) (see e.g., \cite{Peixoto2017}).
Unfortunately, performing these integrals analytically is not possible, because of the structure of the mixture probability.
A numerical solution may exist, but the development of such methods and the assessment of their divergence from the standard non-zero-inflated DCSBM fall outside the scope of this article and warrant dedicated exploration in future research.
\section{Performance of Zero-Inflated Multi-Edge Models}\label{sec:results}
\begin{table}[t]
\centering
\caption{\textbf{Summary of Datasets.} $N$ is the number of nodes, $M$ is the number of unique links (i.e., connected pairs), $m$ is the number of multi-edges, $d$ is the density (fraction of connected pairs), $\rho$ is the multi-edge density (average number of multi-edges per pair of nodes).
Note how all datasets except for BB are very sparse (i.e., $d\ll 1$) despite the large \(\rho\).}\label{tab:data}
\begin{tabular}{lcccccr}
  \hline
Dataset & $N$ & $M$ & $m$ & $d$ & $\rho$ & kurtosis \\ 
  \hline
\textbf{HS13}  & 327 & 5818 & 188508 & 0.11 & 3.54 & 1244.05 \\ 
  \textbf{SFHH}  & 403 & 9565 & 70261 & 0.12 & 0.87 & 4109.62 \\ 
  \textbf{HS12} & 180 & 2220 & 45047 & 0.14 & 2.80 & 712.33 \\ 
  \textbf{WP}  & 92 & 755 & 9827 & 0.18 & 2.35 & 880.08 \\ 
  \textbf{WP15}  & 217 & 4274 & 78249 & 0.18 & 3.34 & 695.74 \\ 
  \textbf{HS11}  & 126 & 1709 & 28561 & 0.22 & 3.63 & 725.30 \\ 
  \textbf{Thiers11}  & 126 & 1709 & 28561 & 0.22 & 3.63 & 725.30 \\ 
  \textbf{LyonSchool}  & 242 & 8317 & 125773 & 0.29 & 4.31 & 237.41 \\ 
  \textbf{HT09}  & 113 & 2196 & 20818 & 0.35 & 3.29 & 1771.89 \\ 
  \textbf{HO}  & 75 & 1139 & 32424 & 0.41 & 11.68 & 152 \\ 
  \textbf{KH}  & 47 & 504 & 32643 & 0.47 & 30.20 & 38.38 \\ 
  \textbf{BB}  & 13 & 78 & 63095 & 1.00 & 808.91 & 10.62 \\ 
   \hline
\end{tabular}\vspace{-2em}
\end{table}

The aim of this section is to investigate the limitations of Poisson multi-edge models in dealing with empirical data and to demonstrate how zero-inflation can address these issues.
We benchmark our models using classical network datasets from the Sociopatterns repository~\cite{sociopatterns}.
These datasets, which report contacts among individuals over short time frames, typically result in sparse multi-edge networks despite a large number of recorded interactions.

The datasets encompass various social interaction scenarios, including interactions among high-school students, conference attendees, and hospital staff.
Each dataset varies in terms of the number of nodes ($N$), unique links ($M$), total multi-edges ($m$), density ($d$), and multi-edge density ($\rho$).
Nevertheless, most datasets exhibit low link density, i.e., $d = M/\binom{N}{2} \ll 1$, and large multi-edge density $\rho = m/\binom{N}{2}$.
This indicates a sparse network structure despite the large number of recorded interactions.
Such characteristics make these datasets a prominent example where classical multi-edge models are sub-optimal.
As shown in \cref{fig:illustration}~(B), naively modelling these datasets would quickly yield fully connected network realisations, in stark contrast with the sparse structure exhibited by the empirical data.

Interestingly, sparsity is often observed together with a ``heavy-taillness'' of the edge count distribution~\cite{Yang2009}.
The empirical distribution of edge counts displays a considerable number of outliers, i.e., unexpectedly large edge counts.
We quantify this by computing the excess kurtosis of the edge count distribution.

Excess kurtosis denotes the tails' heaviness relative to a normal distribution.
Values close to 0 indicate a distribution with similar tail behaviour to the normal distribution.
Positive values signify heavier tails, indicating more extreme outliers, while negative values suggest lighter tails.
In other words, large positive excess kurtosis values imply a higher probability of extreme events or outliers compared to a normal distribution.
The sample excess kurtosis and other basic statistics of the empirical data analysed are reported in \cref{tab:data}.

\subsection{DCSBM versus zi-DCSBM}

\begin{figure}[t]\centering
\includegraphics[width=.8\textwidth]{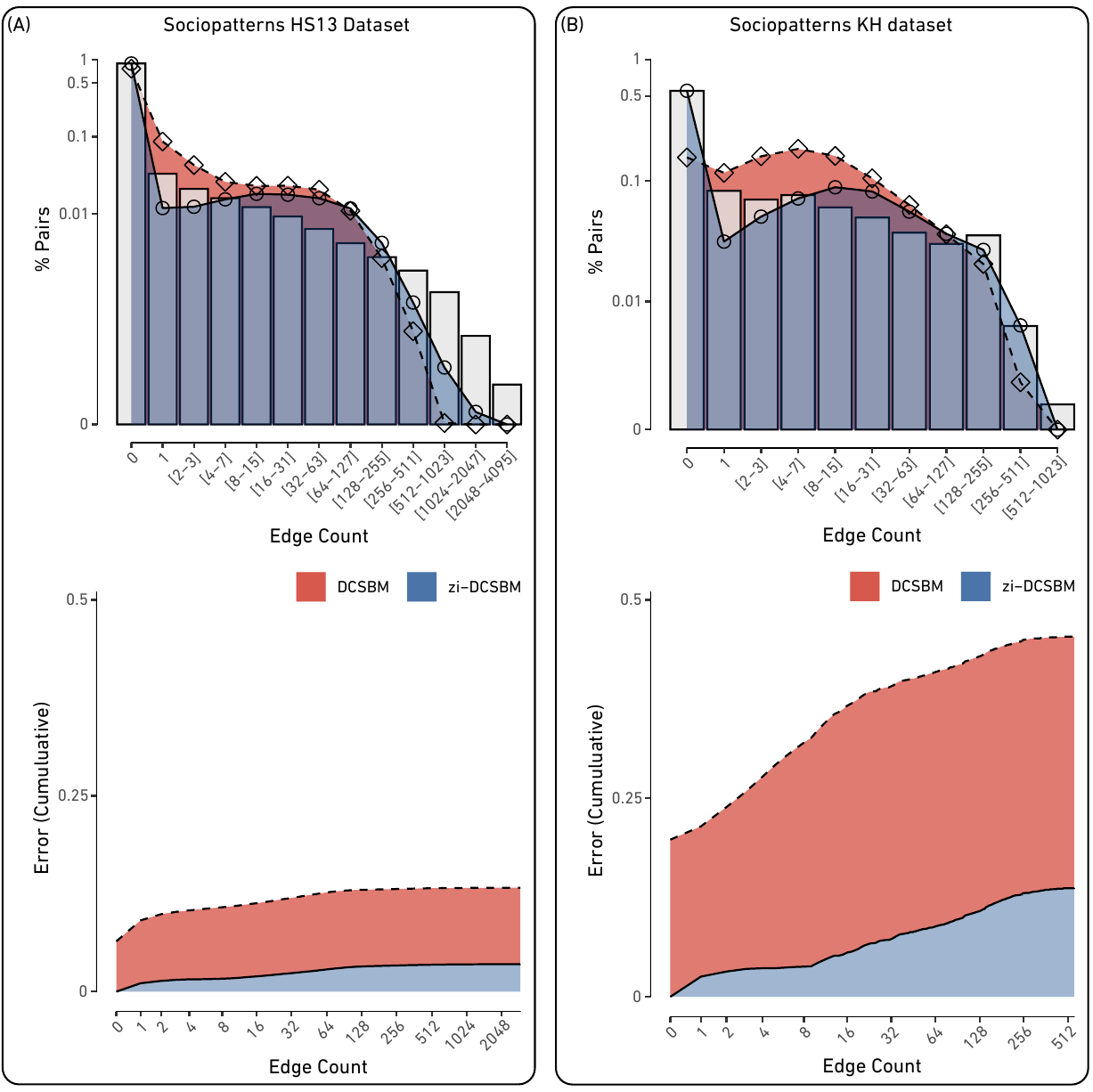}
  \caption{
  	\textbf{(top)} edge count distributions for two exemplary Sociopatterns datasets (HS13 and KH).
  	The grey bar plot shows the empirical edge count distribution.
  	The height of a bar denotes the fraction of pairs in the network connected by a given range of multi-edges.
  	In red, the expected edge count distribution according to a DCSBM whose blocks have been obtained by modularity maximisation.
  	In blue, the expected edge count distribution according to its zero-inflated variant, fitted using the same blocks.
  	\textbf{(bottom)} Cumulative error for the two models.
  	In HS13, most of the difference between the two models can be attributed to the left side of the edge count distribution and pairs with low edge counts.
  	In KH, not only is the DCSBM unable to capture the network sparsity, but it also fails to capture the heavy-tailed nature of the edge count distribution.
  	The zi-DCSBM provides a better fit in both cases.}
  \label{fig:results} 
\end{figure}

Among the models considered in this study, degree-corrected stochastic block models (DCSBM) are the most general due to their capability to accommodate both group-level and node-level heterogeneity.
Consequently, our comparison focuses on the zero-inflated and classical variants of DCSBM.
To maintain simplicity, we opt to infer the blocks in the models utilising the Leiden modularity maximisation algorithm \cite{leiden}.
Modularity, as previously mentioned, relies on the expected edge count derived from an underlying null-model, typically a configuration model.
Given that classical and zero-inflated model pairs share the same expectation, modularity assumes an identical form for both.
Hence, we can utilise identical blocks for both models.
While maximum likelihood optimisation offers the potential for superior models overall, the blocks identified for classical and zero-inflated models are not necessarily the same.
For the purpose of comparing how DCSBM and zi-DCSBM contend with sparse multi-edge networks, employing the same blocks allows us to better understand the differences between the two models.
Therefore, we proceed with modularity-inferred blocks.

\begin{figure}[ht]\centering
\includegraphics[width=.7\textwidth]{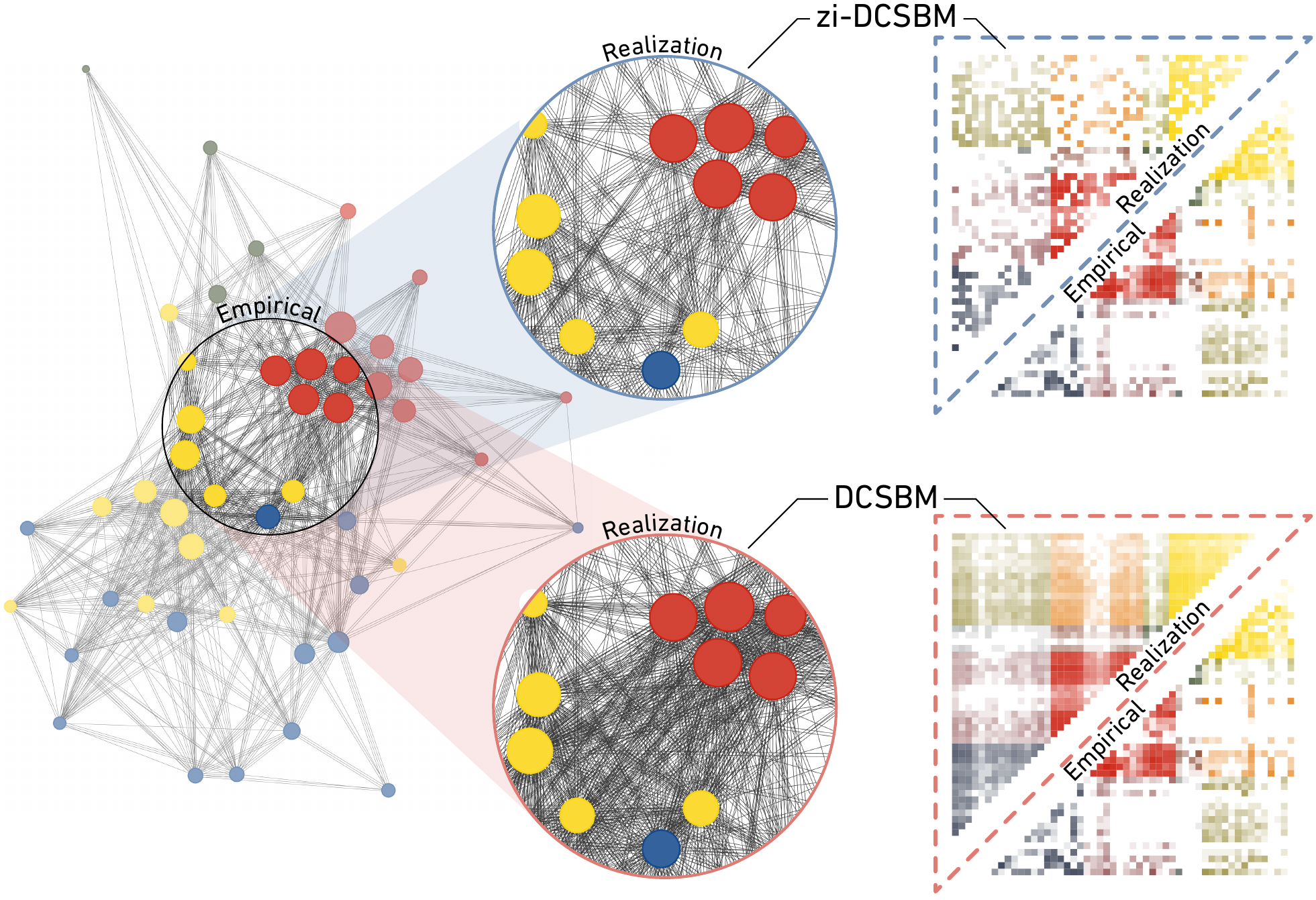}
  \caption{Comparison of the DCSBM and zi-DCSBM fits for the KH dataset.
  On the left side, the network is visualised as a multi-graph with parallel edges denoting multi-edges in log10 base (i.e., 1 edge represents one interaction, 2 parallel edges represent 10 interactions, and so on).
  Nodes are coloured according to the labels inferred by modularity maximisation.
  The ``lens'' plots show a random realisation from the DCSBM (bottom) and zi-DCSBM (top).
  On the right, the adjacency matrices of the random realizations are visualised against the empirical network.
  These plots clearly highlight how the DCSBM fails to capture the sparsity of the empirical data.}
\label{fig:final}
\end{figure}

\paragraph{Sparsity}

When sparsity mainly arises between groups, block models---even without zero-inflation---may appear sufficient to capture sparsity.
However, they are nevertheless outperformed by their zero-inflated variants.
An illustrative example is provided by the HS13 dataset in \cref{fig:results}~(A).
Where the number of multi-edges is large, DCSBM tends to produce numerous connected pairs with low edge counts.
This contrasts with the empirical data, where fewer pairs tend to be connected but with larger edge counts.

In the Supplementary Information, we show that, despite the inherent limitations of the Poisson distribution in capturing heavy tails, introducing zero-inflation redistributes probability mass toward the tails of the edge count distribution, allowing the model to better reflect the extreme values observed in the empirical data.

In \cref{fig:results}~(A) {(top)}, the empirical distribution of edge counts is depicted as a bar plot alongside the expected edge count distribution of DCSBM (shown in red).
It is notable that the bulk of the weight in the distribution of DCSBM is concentrated at small positive counts, thus significantly underrepresenting large counts.
Conversely, the zi-DCSBM, fitted with the same blocks, is capable of shifting the distribution (shown in blue in the plot) towards larger counts.

\Cref{fig:results}~(A) {(bottom)} illustrates the cumulative error over the different edge count values.
It quantifies the discrepancy between the observed distribution and the expected distribution according to the models.
The cumulative error represents the percentage of node pairs with a given edge count in the data compared to those expected in the model.
In the case of HS13, for low counts there is a considerable difference between the DCSBM and the zi-DCSBM.
This difference does not grow further for larger counts.

Additionally, we can use the chi-squared goodness-of-fit statistic to quantitatively compare the distributions.
The chi-squared statistic for the DCSBM is $7199.8$, while the chi-squared statistic for the zi-DCSBM is $4125.2$.
The smaller statistic for the zi-DCSBM shows that the zero-inflated model provides a considerably closer match to the empirical distribution.
Nevertheless, the large value of both statistics indicates that both models deviate significantly from the empirical distribution.

In other examples, multi-edges are bundled over a small number of pairs both within and between groups.
In these cases, the DCSBM performs particularly poorly, as it greatly underestimates the sparsity of the graph.
An example of this is provided by the KH dataset in \cref{fig:results}~(B).
Again, \cref{fig:results}~(B) {(top)} shows the empirical distribution of edge counts as a bar plot and the expected distribution of the DCSBM in red.
Here, the DCSBM is unable to even approximate the empirical distribution.
The zi-DCSBM fitted with the same blocks is instead able to better follow the empirical distribution (in blue in the plot).
Such a large difference can be easily seen in \cref{fig:results}~(B) {(bottom)}.
The cumulative error for the DCSBM starts higher and grows much faster than in the zi-DCSBM case.

These results can be confirmed quantitatively by computing the chi-squared goodness-of-fit statistics.
In the case of the DCSBM, we get $1442.7$.
The zi-DCSBM gives $175.91$, nearly an order of magnitude smaller than the non zero-inflated model.
This supports the qualitative assessment obtained from \cref{fig:results}~(B).
In the supplementary information A% \ref{SI:figs}
, we provide as supplementary figures the equivalents of \cref{fig:results} for the remaining 10 Sociopatterns datasets.

The KH dataset allows for further investigation of the role of zero-inflation on network models.
On the left of \cref{fig:final}, the empirical network is visualised as a multi-edge graph.
The `lens' plots in the centre of the figure show two realisations from the DCSBM (bottom) and zi-DCSBM (top).
It is easy to visually glean what is quantified in \cref{fig:results}.
The DCSBM yields much denser realisations, i.e., with a higher fraction of connected pairs, compared to both its zero-inflated variant and the empirical data.
The adjacency matrices on the right side of \cref{fig:final} further confirm this.
The adjacency matrix of the DCSBM realisation is much denser than either its zero-inflated counterpart or the empirical one.

\begin{figure}[t]\centering
  \includegraphics[width=.9\textwidth]{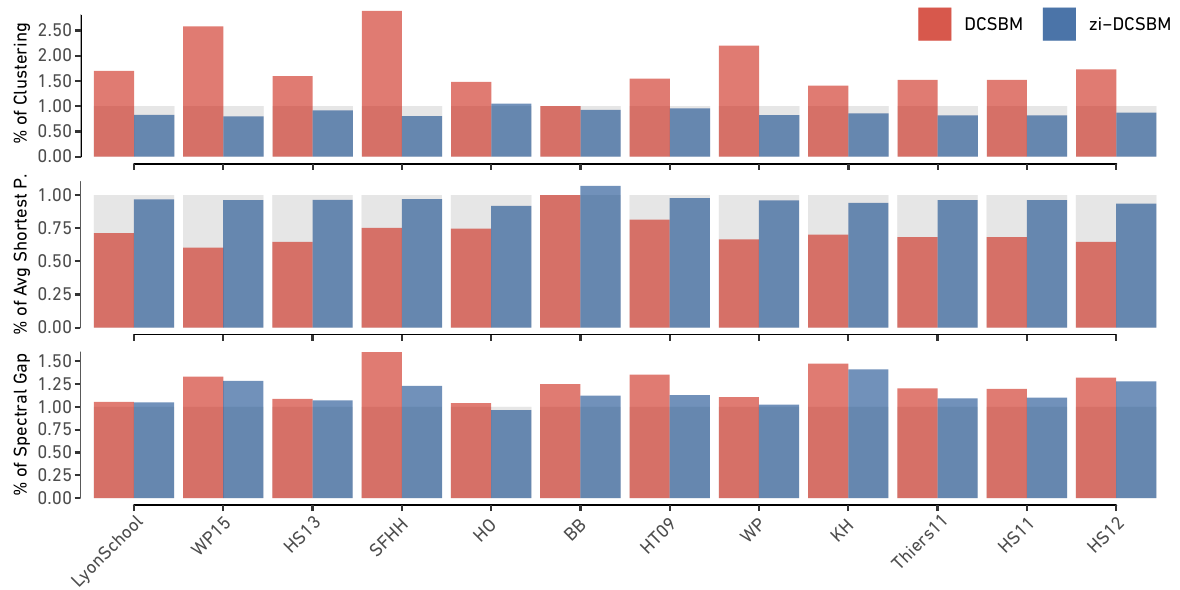}
  \caption{The zi-DCSBM captures properties of empirical networks significantly better that its non-zero inflated variant.
  \textbf{(top)} Percentage of the empirical average clustering coefficient captured by \emph{DCSBM} (red) and \emph{zi-DCSBM} (blue) for all the Sociopatterns datasets.
  \textbf{(center)} Percentage of the empirical average shortest path length captured by \emph{DCSBM} (red) and \emph{zi-DCSBM} (blue) for all the Sociopatterns datasets.
  \textbf{(bottom)} Percentage of the empirical spectral gap captured by \emph{DCSBM} (red) and \emph{zi-DCSBM} (blue) for all the Sociopatterns datasets.
  The expected values of the properties of each model have been computed from 1 000 realisations.}
\label{fig:spectralgap}
\end{figure}

\paragraph{Diffusion Speed}  
The structure of a network significantly impacts dynamics running on it, influencing processes such as information diffusion and opinion formation~\cite{dynamicsbook}.
In denser networks, diffusion processes occur more rapidly.
This can be quantified by the spectral gap--a measure of connectivity and diffusion potential--defined as the difference between the smallest and second smallest eigenvalues of the Laplacian matrix~\cite{Delvenne2015}.
A larger spectral gap generally indicates faster diffusion across the network.

In \cref{fig:spectralgap} (bottom) we show the percentage of the empirical spectral gap captured by the DCSBM and its zero-inflated counterpart across all Sociopatterns datasets.
Our results show that the DCSBM consistently overestimates the diffusion speed, yielding a larger spectral gap than that observed in empirical data.
Conversely, the zi-DCSBM, by better preserving network sparsity, aligns more closely with the observed spectral gap values.
This offers an improved representation of diffusion properties in real-world networks.
In Supplementary Information B, we provide further details about this analysis.

\paragraph{Small-Worldness}
Another key feature distinguishing empirical networks is their small-worldness: 
small world networks are characterized by both a high clustering coefficient and a low average path length.
Small-world networks facilitate efficient communication and diffusion processes, as they maintain short paths between nodes while also forming densely connected clusters.

Due to their higher predicted density, classical network models struggle to reproduce these characteristics.
In \cref{fig:spectralgap} (center), we show the percentage of the empirical average path length captured by the DCSBM and zi-DCSBM models across the datasets, and in \cref{fig:spectralgap}(top), we show the percentage of the empirical clustering coefficient for both models.

Overall, our results demonstrate that the zi-DCSBM provides a substantially closer match to the empirical properties of the network analysed compared to the standard DCSBM.
This improvement highlights the zi-DCSBM’s ability to better preserve network sparsity and realistic connectivity patterns, avoiding the artificial inflation of connectivity that tends to occur in classical multi-edge network models.

\section{Discussion}\label{sec:discussion}

Our study reveals significant limitations of classical multi-edge network models and the potential of zero-inflated models to overcome these challenges.
We have shown that empirical multi-edge networks tend to be sparse despite having a large number of edges.
This sparsity means that many edges are bundled on a few node pairs, resulting in bimodal edge count distributions with excess zeroes and heavy tails.
These characteristics pose considerable difficulties for traditional multi-edge network models as they often fail to capture the sparse nature of real-world networks.
Consequently, key features of empirical networks such as diffusion speed and small-worldness are misrepresented.
Hence, the practical usability of these network models is limited.

To mitigate these limitations, we show how classical multi-edge network models can be extended to incorporate zero-inflation.
This mechanism accounts for the excess number of zeroes (disconnected pairs) observed in empirical data.
Zero-inflation not only helps reproduce network sparsity but also remedies several critical shortcomings of classical multi-edge network models:
Both diffusion speed and small-worldness are better captured via zero-inflated models.

Our results indicate that, while zero-inflation significantly enhances the fit of the models to empirical data, there remain areas for improvement.
As shown in Supplementary Information D, one important challenge is computational complexity:
the numerical maximum-likelihood estimation scales poorly with large networks, especially when considering heterogeneous zero-inflation.
Moreover, in this article, we have used modularity maximization to infer ``model-independent'' node labels, allowing us to compare zero-inflated models with classical counterparts without the confounding effects of differing block structures.
However, developing block inference algorithms specifically tailored for zero-inflated models would yield more accurate representations of network structures.
This is particularly important because blocks derived from modularity maximization, while useful, do not always capture the full complexity of empirical networks and do not fully exploit the advantages of zero-inflation.

Furthermore, the zero-inflated Poisson models used here are limited by the restrictive mean-variance relationship of the Poisson distribution, which may not adequately account for over-dispersion—a common feature of count data~\cite{Yang2009}.
Employing more flexible distributions, such as the negative binomial or generalized hypergeometric distributions, could improve model fit by better capturing over-dispersion, particularly in networks with more complex count structures~\cite{choi2023model, Dong2019}.

Our findings suggest that zero-inflated models provide a more detailed understanding of networks, capturing both the sparsity and the extreme events reflected in the data.
This aligns with the need for accurate models in various applications, such as optimising distribution systems, understanding disease spread, and analysing social behaviours.
In our study, the example networks are so sparse that the necessity of zero-inflation is evident.
However, in less extreme cases, determining the need for zero-inflation versus an appropriate choice of block structure becomes essential.
To address this, appropriate likelihood-ratio tests and model comparison techniques have been developed for various models~\cite{choi2023model, Motalebi2021dcsbm, Dong2019, casiraghimle}.
These methods can help ascertain the necessity of zero-inflation by comparing the fit and performance of different models on the same dataset, providing a rigorous basis for model selection.
Further work in this direction is needed to adapt these model selection techniques to high-dimensional, sparse network settings where zero-inflation may play a critical role in accurately capturing network structure.

\section*{Aknowledgements}
The authors thank Prof. Frank Schweitzer for his support and Dr. Giacomo Vaccario for useful discussions.
G.A. acknowledges funding from SNF Grant n.192746.

% ---- Bibliography ----
%
{\small \setlength{\bibsep}{1pt}

}

\end{document}

\appendix
\renewcommand{\thefigure}{S\arabic{figure}}
\title{Supplementary Information -- \\\normalsize Empirical Networks are Sparse: Enhancing Multi-Edge Models with Zero-Inflation}
\titlealternative{SI -- Empirical Networks are Sparse: Enhancing Multi-Edge Models with Zero-Inflation}
\maketitle
\renewcommand*{\thefootnote}{\arabic{footnote}}
\section{Supplementary Figures}\label{SI:figs}
\begin{figure}[ht]\centering
\fbox{
  \includegraphics[width=.45\textwidth]{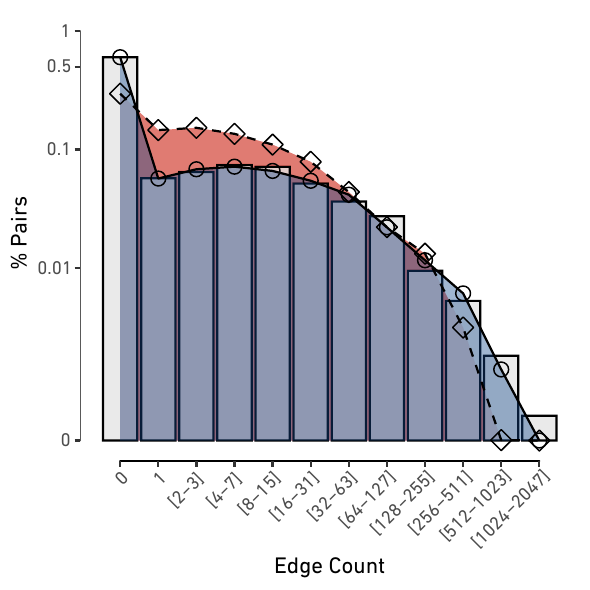}
  \includegraphics[width=.45\textwidth]{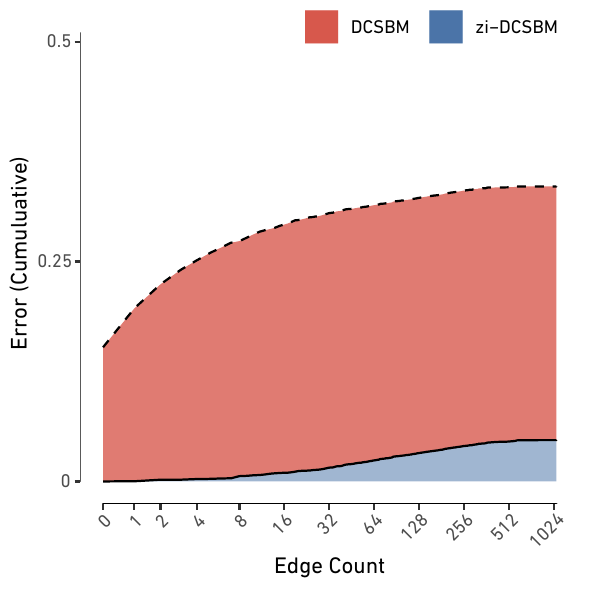}
}
\caption{Edge count distribution and errors for the HO dataset.}

\end{figure}
\begin{figure}[ht]\centering
\fbox{
  \includegraphics[width=.45\textwidth]{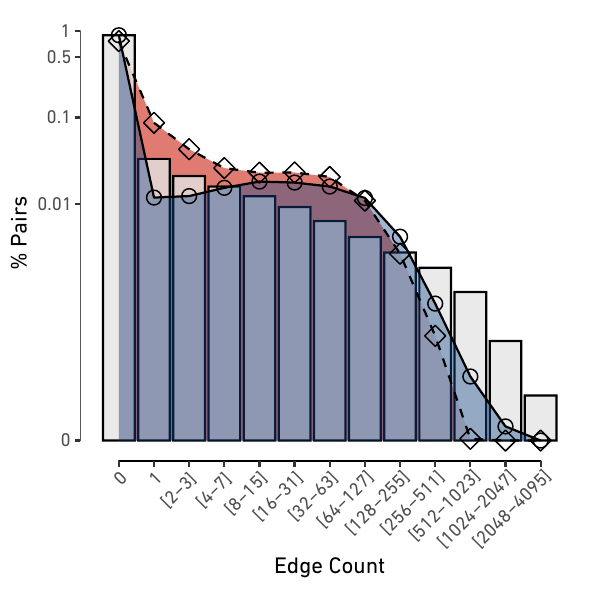}
  \includegraphics[width=.45\textwidth]{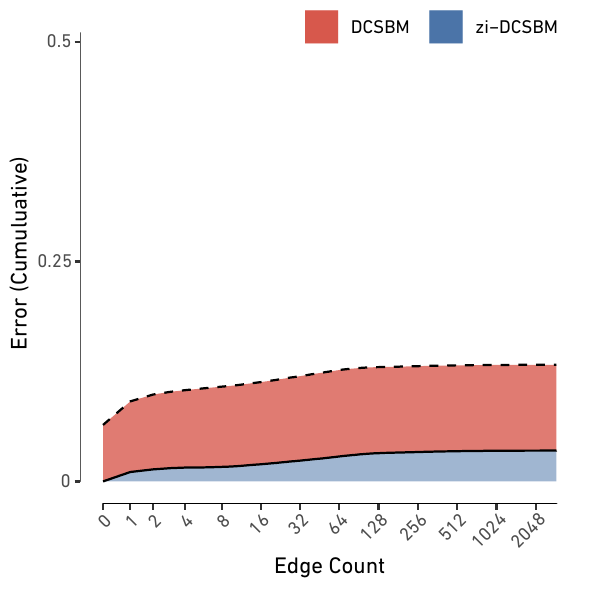}
}
\caption{Edge count distribution and errors for the HS13 dataset.}

\end{figure}                          
\begin{figure}[ht]\centering
\fbox{
  \includegraphics[width=.45\textwidth]{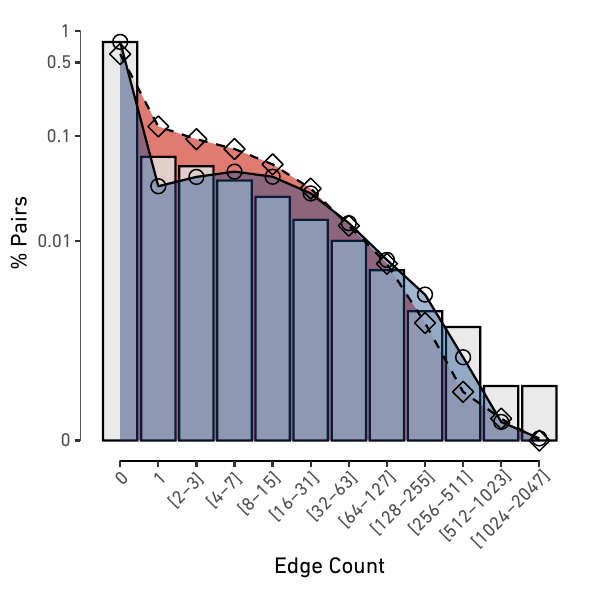}
  \includegraphics[width=.45\textwidth]{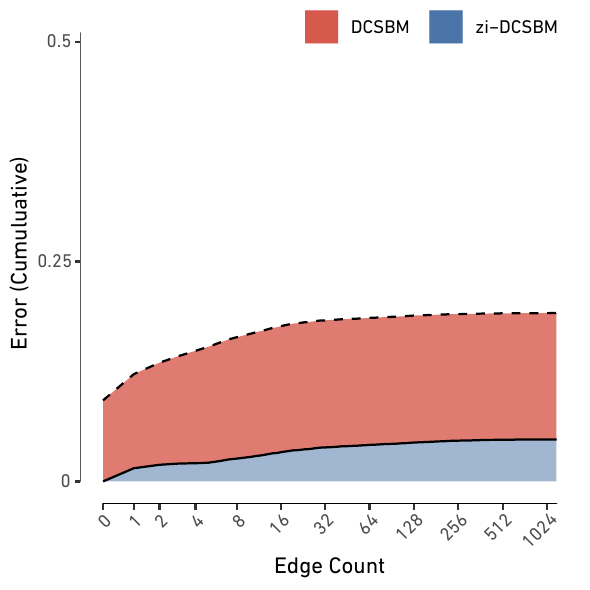}
}
\caption{Edge count distribution and errors for the HS11 dataset.}

\end{figure}                                      
\begin{figure}[ht]\centering
\fbox{
  \includegraphics[width=.45\textwidth]{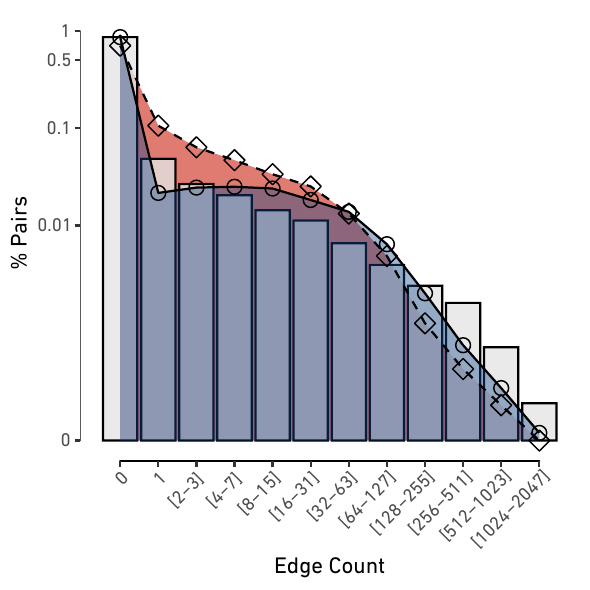}
  \includegraphics[width=.45\textwidth]{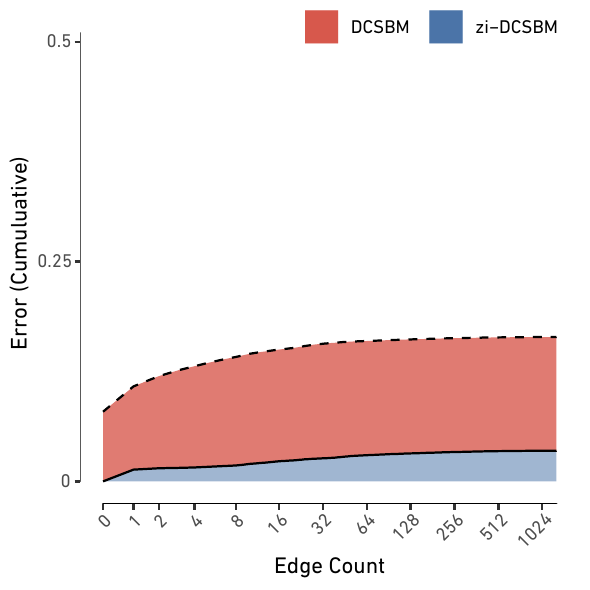}
}
\caption{Edge count distribution and errors for the HS12 dataset.}

\end{figure}                                      
\begin{figure}[ht]\centering
\fbox{
  \includegraphics[width=.45\textwidth]{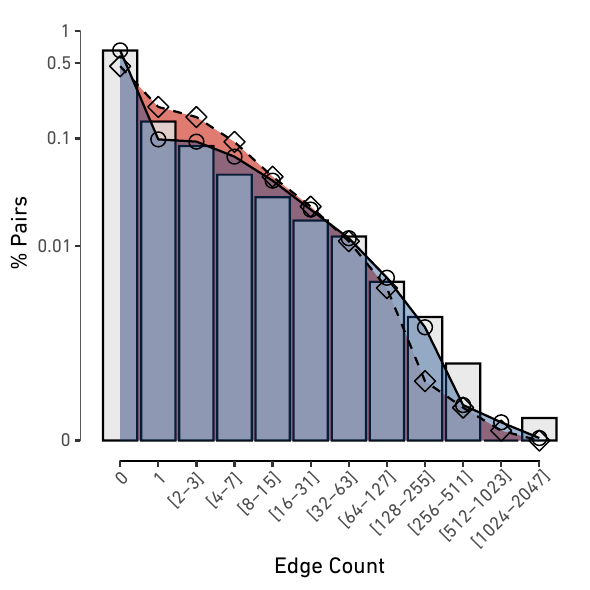}
  \includegraphics[width=.45\textwidth]{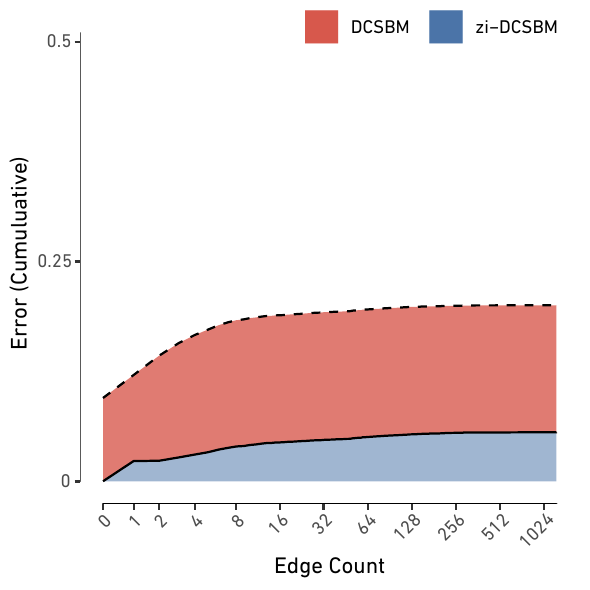}
}
\caption{Edge count distribution and errors for the HT09 dataset.}

\end{figure}                                  
\begin{figure}[ht]\centering
\fbox{
  \includegraphics[width=.45\textwidth]{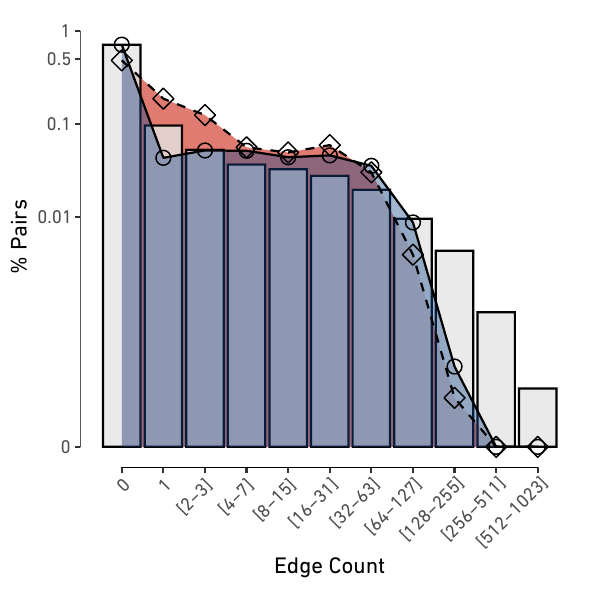}
  \includegraphics[width=.45\textwidth]{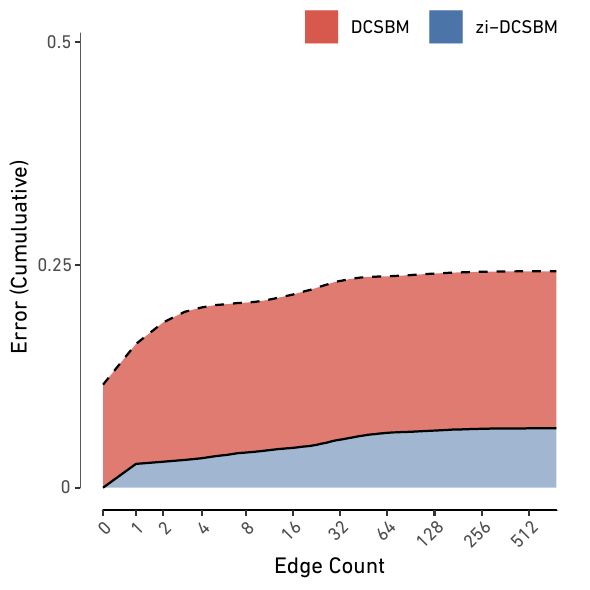}
}
\caption{Edge count distribution and errors for the LyonSchool dataset.}

\end{figure}                                          
\begin{figure}[ht]\centering
\fbox{
  \includegraphics[width=.45\textwidth]{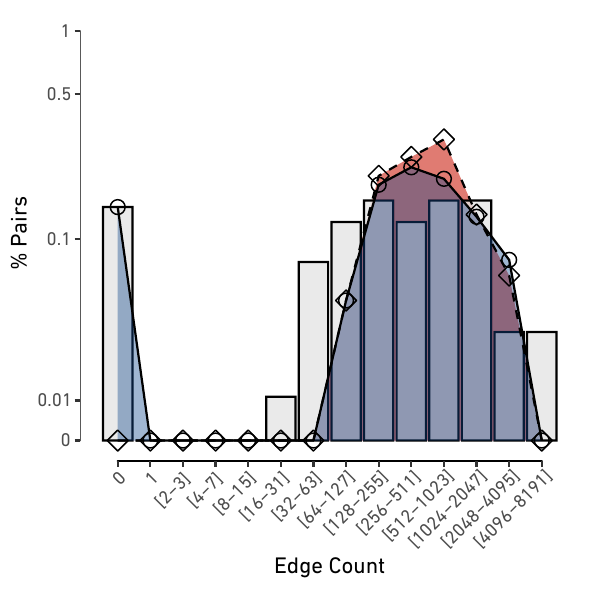}
  \includegraphics[width=.45\textwidth]{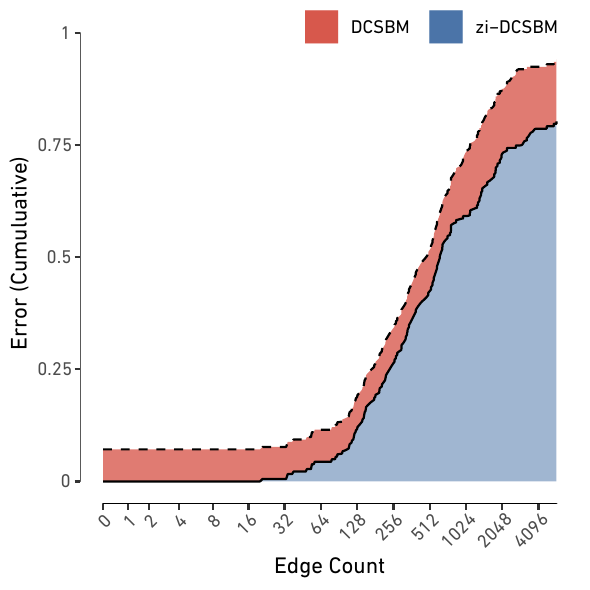}
}
\caption{Edge count distribution and errors for the BB dataset.}

\end{figure}                                                  
\begin{figure}[ht]\centering
\fbox{
  \includegraphics[width=.45\textwidth]{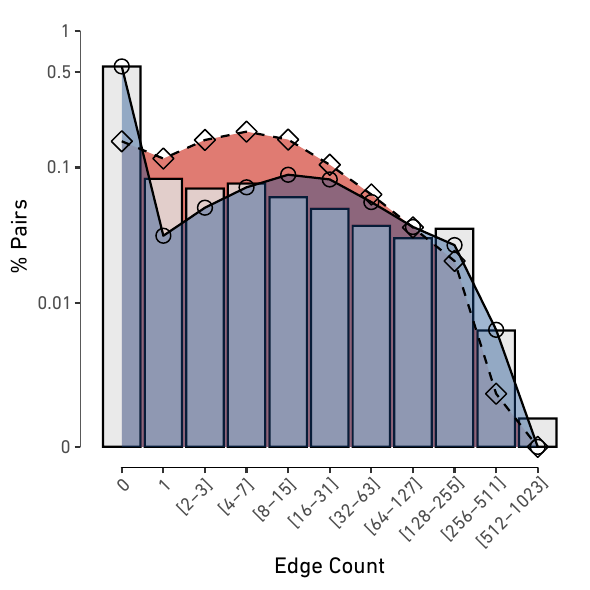}
  \includegraphics[width=.45\textwidth]{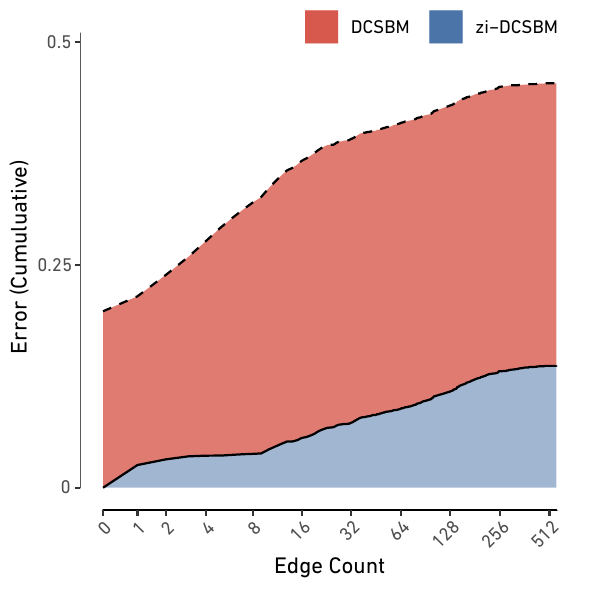}
}
\caption{Edge count distribution and errors for the KH dataset.}

\end{figure}                  
\begin{figure}[ht]\centering
\fbox{
  \includegraphics[width=.45\textwidth]{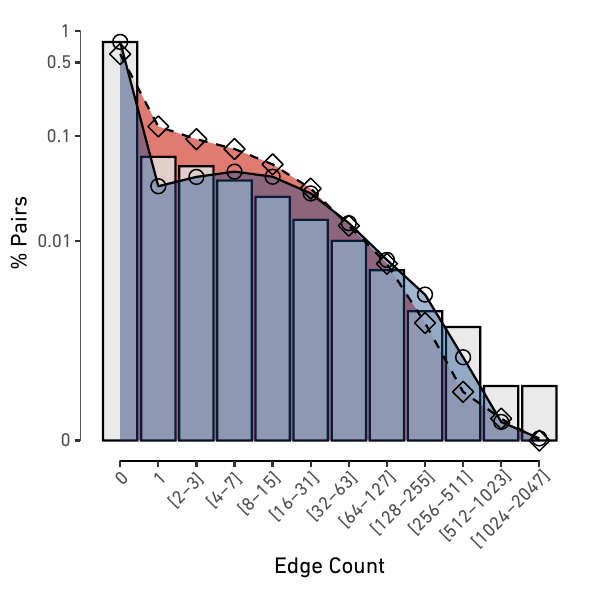}
  \includegraphics[width=.45\textwidth]{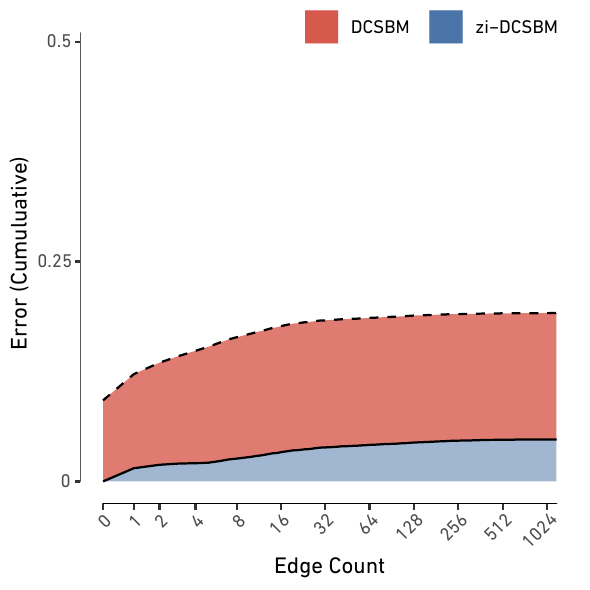}
}
\caption{Edge count distribution and errors for the Thiers11 dataset.}

\end{figure}                                              
\begin{figure}[ht]\centering
\fbox{
  \includegraphics[width=.45\textwidth]{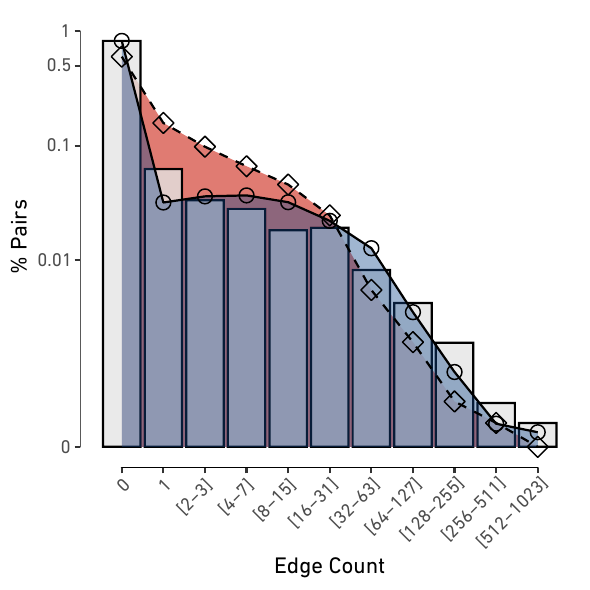}
  \includegraphics[width=.45\textwidth]{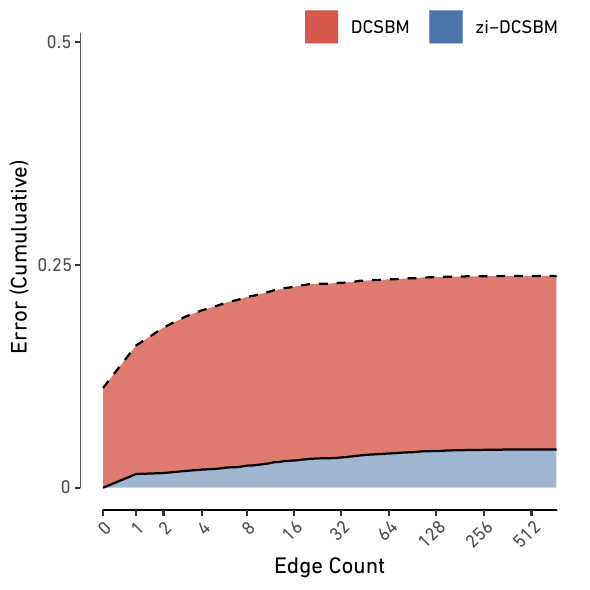}
}
\caption{Edge count distribution and errors for the WP dataset.}

\end{figure}                                                    
\begin{figure}[ht]\centering
\fbox{
  \includegraphics[width=.45\textwidth]{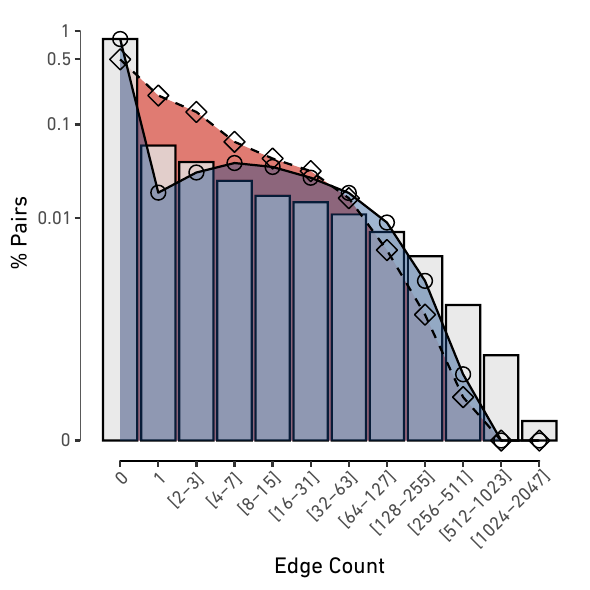}
  \includegraphics[width=.45\textwidth]{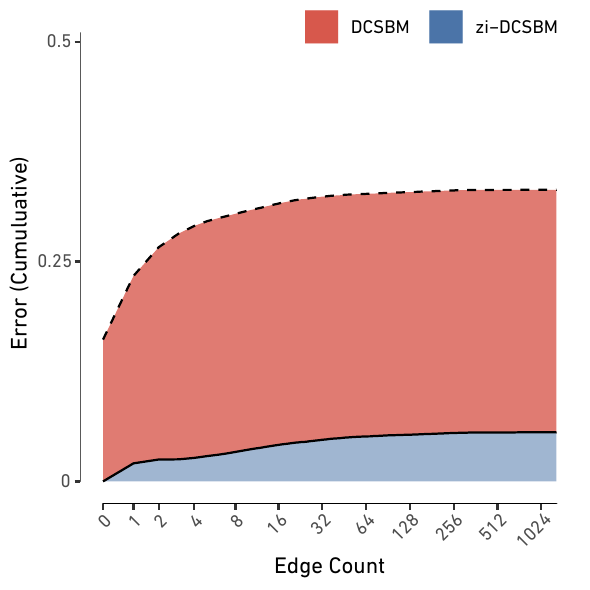}
}
\caption{Edge count distribution and errors for the WP15 dataset.}

\end{figure}                                                
\begin{figure}[ht]\centering
\fbox{
  \includegraphics[width=.45\textwidth]{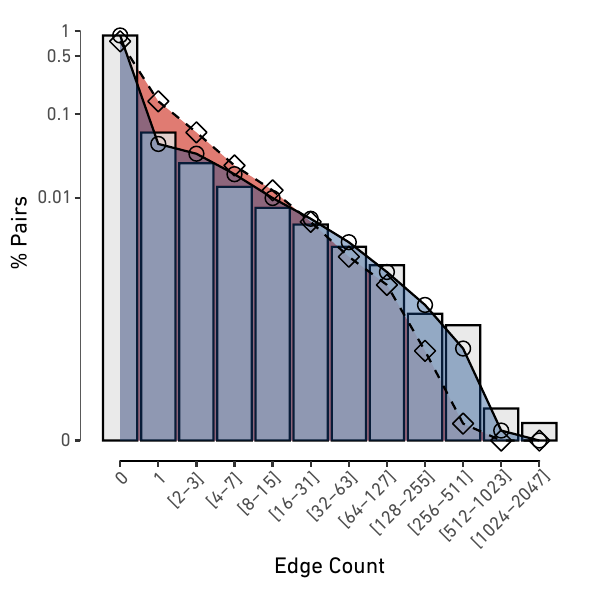}
  \includegraphics[width=.45\textwidth]{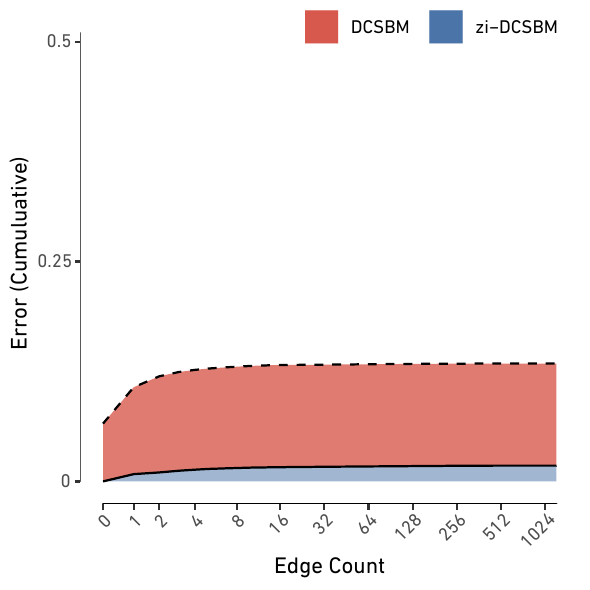}
}
\caption{Edge count distribution and errors for the SFHH dataset.}

\end{figure}

\clearpage
\section{Diffusion Speed}\label{SI:spectral}

We investigate how sparsity influences the diffusion speed of dynamical processes running on networks.
Diffusion speed is crucial for understanding various phenomena, such as information diffusion and opinion formation~\cite{dynamicsbook}.
One way to estimate the diffusion speed is by computing the spectral gap of the network~\cite{Delvenne2015}.

The spectral gap is defined as the difference between the smallest (which is always zero for connected graphs) and the second smallest eigenvalue of the laplacian matrix of a graph~\cite{chung1997spectral}.
This gap provides insights into the connectivity and overall structure of the network.
A larger spectral gap generally indicates faster diffusion, as it reflects higher connectivity and fewer bottlenecks in the network~\cite{Delvenne2015}.
The laplacian matrix $L$ of a graph is given by
\begin{equation}
  L = I - D^{-1} T,
\end{equation}
where $I$ is the identity matrix, $D$ is the diagonal matrix built from the vector of degrees, and $T$ is the transition matrix obtained by row-normalising the adjacency matrix of a multi-edge graph.

Here, we compute the spectral gap of all the networks in the Sociopatterns repository.
Similarly to the analysis performed for excess kurtosis in Section 3.1
in the main text, we further estimate the expected spectral gap according to the DCSBM and zi-DCSBM for each dataset.

Our results show that DCSBM consistently overestimates the diffusion speed in all datasets.
The average spectral gap from the DCSBM is significantly larger than that computed from the empirical data.
A student's t-test performed on 1000 realisations shows a highly significant difference (p-value < 1e-16) in all datasets.
The expected spectral gap from the zi-DCSBM is significantly lower than that from the non-zero-inflated model (all p-values < 1e-7), indicating a closer match to the empirical data.

\section{Excess Kurtosis}\label{SI:kurtosis}
\begin{figure}[t]\centering
  \includegraphics[width=.8\textwidth]{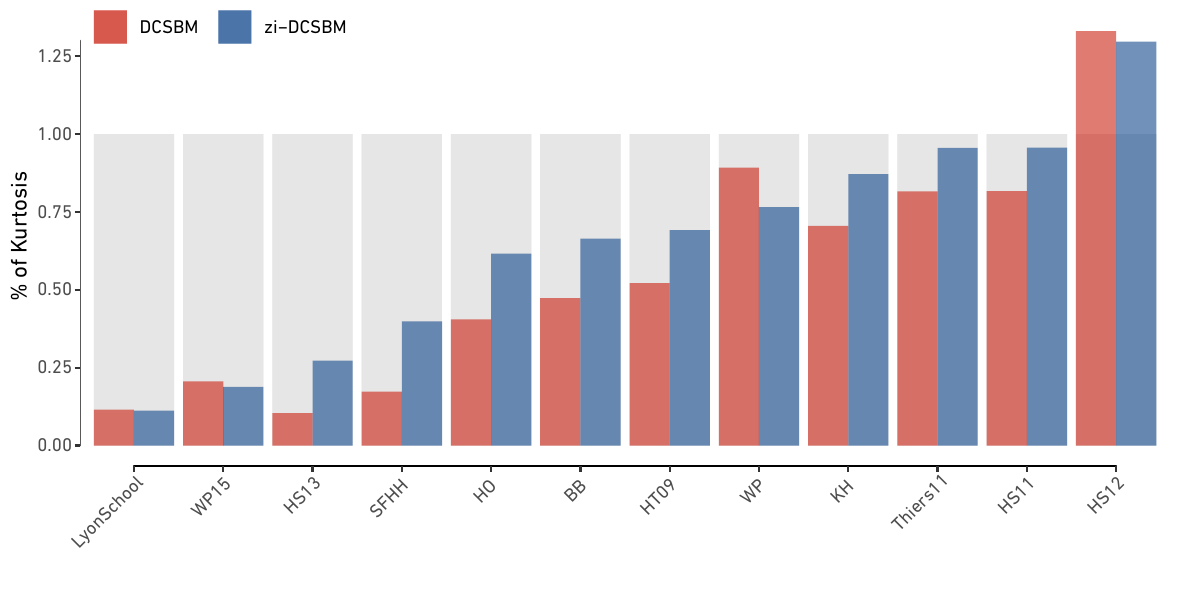}
  \caption{Percentage of \emph{sample} excess kurtosis of the edge count distribution captured by \emph{DCSBM} (red) and \emph{zi-DCSBM} (blue) for all the Sociopatterns datasets. The expected kurtosis of each model has been computed from 100 000 realisations. Generally, zero-inflated model variants shift part of the weight of the edge count distribution to the tail, increasing the kurtosis and more closely following the empirical data.}
\label{fig:kurtosis}
\end{figure}

 In \cref{fig:kurtosis}, we plot the percentage of sample excess kurtosis in the edge count distribution captured by the DCSBM (red) and zi-DCSBM (blue). This plot allows us to compare the ability of the two models to capture excess kurtosis in real data.

Zero-inflated model variants, such as the zi-DCSBM, shift part of the probability mass from the center of the edge count distribution toward the tails. This adjustment (i) increases kurtosis, thus better reflecting the heavy tails often observed in empirical data, and (ii) improves the overall fit to the data by aligning more closely with the distribution's shape. As shown in the plot, zero-inflated models tend to yield higher kurtosis where the empirical data exhibits high sample kurtosis, indicating an ability to capture more extreme edge counts than standard model variants.

However, while zero-inflated Poisson models offer an improvement over traditional Poisson models, they are limited by the fixed relationship between mean and variance inherent to the Poisson distribution. This constraint restricts their capacity to fully capture the heavy tails in the data. Using alternative distributions, such as the negative binomial, could provide further flexibility by introducing additional parameters to disentangle the mean and variance, thereby enhancing the fit to highly variable network data. As discussed, we focus here on zero-inflated Poisson models due to their interpretability and natural extension of classical network models, but we recognize that exploring other distributions for edge counts represents a promising direction for future research.

\section{Computational Resources}

To evaluate the computational resources required for estimating the zero-inflated Degree-Corrected Stochastic Block Model (zi-DCSBM), we analyzed the time taken for parameter estimation across various configurations of node and block counts. Our results indicate that the estimation time depends primarily on the number of zero-inflation parameters. While the time required does increase with the number of nodes, it grows more dramatically with the number of blocks.

\paragraph{Experimental Setup}

For each configuration, we generated 100 realizations of random multi-edge graphs with a specified density and multi-edge structure. Specifically, we generated networks with random block structures but given number of blocks, fixed density, and a specified number of multi-edges. For simplicity, we kept the number of multi-edges at \(10 N^2\) and sampled the density uniformly at random between 0 and 0.5, as we found that these parameters did not impact significantly the estimation time. The number of nodes \(N\) was varied between 10 and 200, and the number of blocks \(B\) between 2 and 20.

\paragraph{Scaling of Computational Complexity}
The zi-DCSBM model, given by

\begin{align}\label{eq:zidcsbm_si}
    \begin{split}
  \Pr(\mathcal{G}|\pmb{\theta},\pmb{\lambda},\pmb{q}) &= \prod_{ij} \biggl( (1-q_{b_i b_j})\,\delta_0(A_{ij}) +
                                                      q_{b_i b_j}\,\frac{(\theta^\text{out}_i \theta^\text{in}_j \lambda_{b_i b_j})^{A_{ij}}}{A_{ij}!}\exp(-\theta^\text{out}_i \theta^\text{in}_j \lambda_{b_i b_j}) \biggr),
    \end{split}
\end{align}
is characterized by Poisson parameters \(\pmb{\theta}\) and \(\pmb{\lambda}\) for the block structure and zero-inflation parameters \(\pmb{q}\) for zero-inflation.

The primary reason for the faster growth in computational complexity with the number of blocks, rather than nodes, is related to the distinct estimation processes for the parameters. Thanks to the fact that we can use the first moment equation to directly tie to observed quantities the Poisson parameters \(\pmb{\theta}\) and \(\pmb{\lambda}\), their estimation is straightforward and requires only simple assignments. However, the zero-inflation parameters \(\pmb{q}\) must be estimated by optimizing the likelihood function.

In a directed zi-DCSBM model, the likelihood function can be decomposed into \(B^2\) independent optimization problems, one for each \(q_{b d}\). Consequently, the time needed to estimate these parameters increases significantly with the number of blocks, as each additional block introduces more zero-inflation parameters to estimate. Although the independence of these optimization problems would allow parallel estimation, the results shown in the figure are computed using a single core.

\begin{figure}[t]\centering
  \includegraphics[width=.8\textwidth]{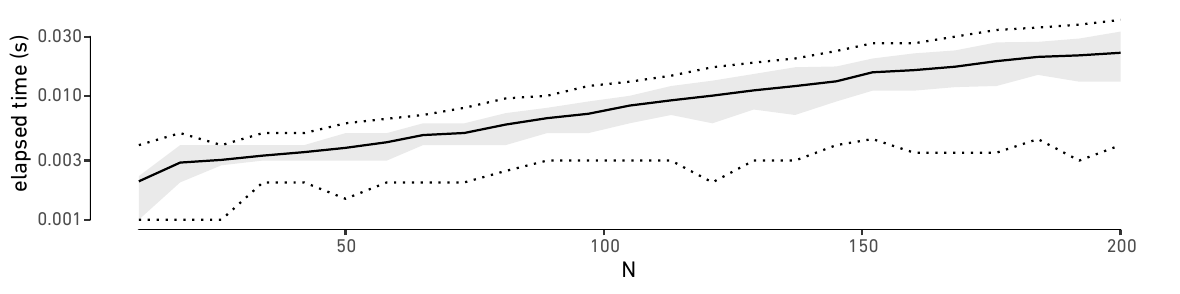}\\
    \includegraphics[width=.8\textwidth]{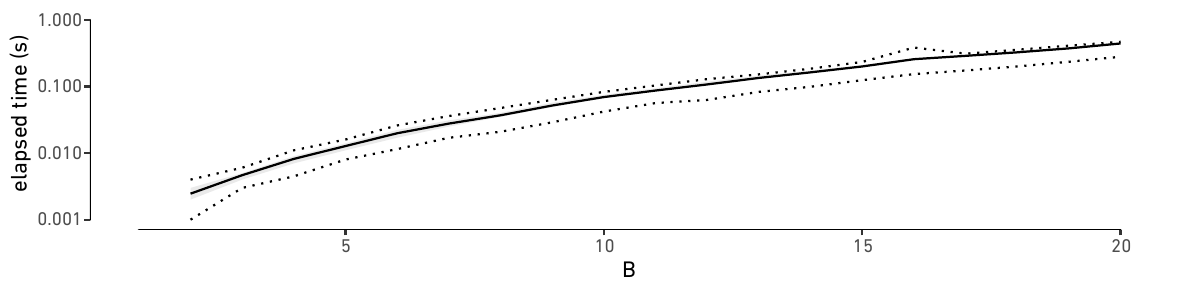}
  \caption{Average time (solid line) needed for parameter estimation of a zero-inflated degree-corrected stochastic block model as a function of the number of nodes $N$ \textbf{(top)} and the number of blocks $B$ \textbf{(bottom)}. The dotted lines denote the 95\% quantiles and the shaded area the interquartile range for the parameter estimation of 100 random networks per values of $N$ and $B$.}
\label{fig:complexity}
\end{figure}

\paragraph{Extended Experiment with Node-Level Zero-Inflation}

To further assess how computational requirements scale with the number of zero-inflation parameters, we conducted a second experiment with a zero-inflated configuration model (zi-CLCM) that includes node-level zero-inflation. In this model, each node \(i\) is assigned a Poisson parameter \(\theta_i\) and a zero-inflation parameter \(q_i\), resulting in the following distribution:

\begin{align}\label{eq:ziclcm_si}
    \begin{split}
      \Pr&(\mathcal{G}|\pmb\theta,\pmb q) = \prod_{ij} \left( (1-q_i q_j)\,\delta_0(A_{ij}) + q_i q_j\,\frac{(\theta^\text{out}_i \theta^\text{in}_j)^{A_{ij}}}{A_{ij}!}\exp(-\theta^\text{out}_i \theta^\text{in}_j) \right).
    \end{split}
\end{align}

Also for this model, the Poisson parameters \(\pmb{\theta}\) can be estimated using the first moment equation:

\[
\widehat{\theta^\star_i} = \frac{\widehat{k^\star_i}}{\sqrt{\widehat{m} q_i}}.
\]

\begin{figure}[t]\centering
  \includegraphics[width=.8\textwidth]{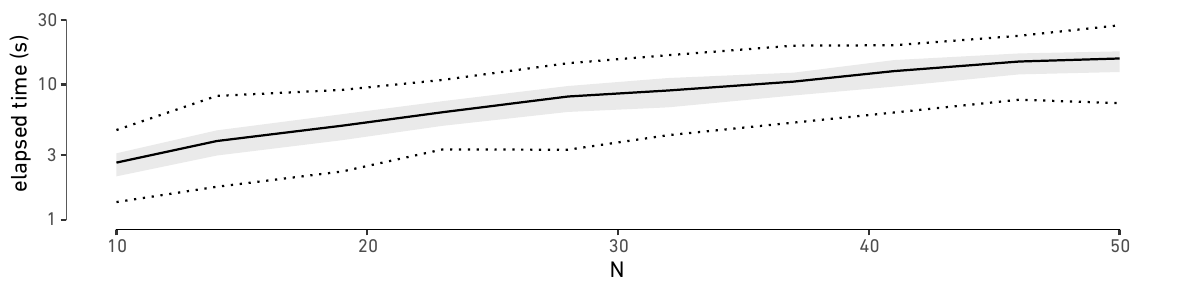}
  \caption{Average time (solid line) needed for parameter estimation of a zero-inflated configuration model with node-level zero-inflation as a function of the number of nodes $N$. The dotted lines denote the 95\% quantiles and the shaded area the interquartile range for the parameter estimation of 100 random networks per value of $N$.}
\label{fig:complexity_clcm}
\end{figure}

However, unlike the previous example, the likelihood function here cannot be decomposed into independent optimization problems, as each node’s zero-inflation parameter \(q_i\) interacts with all others. As a result, the time required for parameter estimation now increases considerably with the number of nodes, as shown in the \cref{fig:complexity_clcm}.

\paragraph{Summary}

Overall, our experiments highlight that the computational cost of estimating zero-inflated models is largely driven by the number of zero-inflation parameters, with a substantial increase in complexity they grow. Future work could explore more efficient strategies to reduce estimation time, especially in cases where node-level zero-inflation parameters are required.

\paragraph{Computational Specifications}

All computations were performed on a machine with the following specifications:

\begin{itemize}
    \item \textbf{CPU:} Intel(R) Core(TM) i9-7960X @ 2.80GHz, 16 cores, 32 threads
    \item \textbf{Max CPU Frequency:} 4.4 GHz
    \item \textbf{L1 Cache:} 32 KB (per core)
    \item \textbf{L2 Cache:} 1 MB (per core)
    \item \textbf{L3 Cache:} 22.5 MB (shared)
    \item \textbf{Total CPU(s):} 32 (hyper-threaded)
\end{itemize}

Unless otherwise specified, all parameter estimations were conducted on a single core.

\section{Reparametrizing zi-DCSBM}

In the zi-DCSBM, we require a set of constraints for identifiability, as the model parameters \(\pmb{\theta}^{\star}\) are defined modulo constants within each block:
\begin{equation}
  \label{eq:dcsbm-constraint_si}
  \sum_{i}\theta^\star_i\delta_b(b_i) = C_b^\star.
\end{equation}

In the following, we show that any choice of value for the parameters $C^\star_{b}$ in the zi-DCSBM is possible since the extra parameter is absorbed by the block parameters $\lambda_{bd}$.
From the constraint in \cref{eq:dcsbm-constraint_si} and Eq. (28) in the main manuscript, we get
\begin{equation}
  \label{eq:sup lambda}
  \widehat{\lambda}_{bd} = \frac{\widehat{m}_{bd}}{q_{bd}C_b^{\text{out}} C_d^{\text{in}}}.
\end{equation}
Then, from the moment equations in Eq. (30) in the main manuscript, we find
\begin{equation}
  \label{eq:sup k}
  \mathbb{E}(k^{\text{out}}_{i}|\pmb{\theta}, \pmb{q}) = \theta_{i}^{\text{out}} \sum_{d} C_d^{\text{in}} q_{b_{i}d} \lambda_{b_{i}d}, \qquad 
  \mathbb{E}(k^{\text{in}}_{i}|\pmb{\theta}, \pmb{q}) = \theta_{i}^{\text{in}} \sum_{d} C_d^{\text{out}} q_{b_{i}d} \lambda_{b_{i}d}.
\end{equation}
By replacing \cref{eq:sup lambda} into \cref{eq:sup k}, we find expressions for the estimators of $\pmb{\theta^{\text{out}}}$ and $\pmb{\theta^{\text{in}}}$:
\begin{equation}
  \label{eq:sup theta}
  \widehat{\theta^{\star}_{i}} = C^\star_{b_i}\frac{\widehat{k^{\star}_{i}}}{\widehat{\kappa}^{\star}_{b_{i}}}.
\end{equation}
The Poisson rate for each pair $(i,j)$ in the full model, Eq. (26), is given by $\theta^{\text{out}}_{i}\theta^{\text{in}}_{j}\lambda_{b_{i}b_{j}}$.
By substituting the MLE's derived in \cref{eq:sup lambda,eq:sup theta}, it becomes apparent that the full model does not depend on the parameters $C^\star_{b}$.

{\small \setlength{\bibsep}{1pt}

}